\documentclass{jfm}

\usepackage{graphicx}
\usepackage{newtxtext}
\usepackage{newtxmath}
\usepackage{natbib}
\usepackage{hyperref}
\usepackage{optidef}
\usepackage{algorithm}
\usepackage{algorithmic}

\hypersetup{
    colorlinks = true,
    urlcolor   = blue,
    citecolor  = black,
}

\newcommand{\RomanNumeralCaps}[1]
\linenumbers

\title{Extracting self-similarity from data}

\author{Nikos Bempedelis\aff{1}
  \corresp{\email{n.bempedelis@qmul.ac.uk}},
  Luca Magri\aff{2,3,4}
 \and Konstantinos Steiros\aff{2} \corresp{\email{k.steiros@imperial.ac.uk}}}

\affiliation{\aff{1}School of Engineering and Materials Science, Queen Mary University of London, E1 4NS, London, UK
\aff{2}Department of Aeronautics, Imperial College London, SW7 2AZ London, UK
\aff{3}The Alan Turing Institute, London NW1 2DB, UK
\aff{4}Politecnico di Torino, DIMEAS, Corso Duca degli Abruzzi, 24 10129 Torino, Italy
}

\begin{document}
\maketitle

\begin{abstract}
Identifying self-similarity is key to understanding and modelling a plethora of phenomena in fluid mechanics. Unfortunately, this is not always possible to perform formally in highly complex flows. We propose a methodology to extract the similarity variables of a self-similar physical process directly from data, without prior knowledge of the governing equations or boundary conditions, based on an optimization problem and symbolic regression. We analyze the accuracy and robustness of our method in five problems which have been influential in fluid mechanics research: a laminar boundary layer, Burger's equation, a turbulent wake, a collapsing cavity, and decaying turbulence. Our analysis considers datasets acquired via both numerical and wind tunnel experiments. The algorithm recovers the known self-similarity expressions in the first four problems and generates new insights on single length scale theories of homogeneous turbulence. 
\end{abstract}


\section{Introduction}
The various constraints that describe a physical phenomenon are often reflections of underlying symmetry principles, summarising regularities that exist independent of specific dynamics \citep{gross1996role}. A notable example is the constraint of momentum conservation, which reflects a translational symmetry of the Euler-Lagrange equations \citep{Landau1976}. Dimensional analysis, historically linked to the discovery of scaling laws and non-dimensional numbers \citep{barenblatt1996scaling,cantwell2002introduction}, expresses the principle of covariance, i.e., the symmetry of any physical law under a dilational transformation of its units of measurement. More generally, when one considers a specific problem, i.e., the governing physical laws and the corresponding boundary and/or initial conditions, multiple symmetries may be simultaneously present (e.g., spiral, reflectional, rotational) \citep{pakdemirli1998similarity,cantwell2002introduction}.

A type of symmetry which is of fundamental importance across various branches of physics is that of self-similarity \citep{barenblatt1996scaling,cantwell2002introduction,tennekes1972first,townsend1976structure}. Following \citet{barenblatt1996scaling,townsend1976structure,pope_2000}, we refer to self-similar (or self-preserving) phenomena as those whose evolution remains invariant under the group transformations of dilation (e.g., heat diffusion), translation (e.g., travelling waves), or a combination of both (e.g., turbulent wakes). Self-similarity allows for the reduction of an $n-$independent variable partial differential equation system into a system with $n-1$ \textit{similarity variables} \citep{pakdemirli1998similarity,cantwell2002introduction,birkhoff2015hydrodynamics}. The transformation from original to similarity variables is known as a similarity transformation. Similarity transformations are particularly useful in problems involving $n=2$ independent variables because they transform a partial differential equation into an ordinary differential equation, which is more amenable to be analytically solved \citep{pakdemirli1998similarity,cantwell2002introduction}. Even if the governing equations are unknown, similarity considerations allow for the grouping of the problem parameters into similarity variables, reducing the effort required by the experimentalist when performing parametric characterizations of a problem by orders of magnitude \citep{barenblatt1996scaling}. Additionally, the identification of similarity variables, subject to the constraints of the governing equations, can lead to the derivation of scaling laws, which describe the asymptotic evolution of the problem variables \citep{tennekes1972first,townsend1976structure,beaumard2024scale}. The identification of similarity transformations and variables has played a prominent role in fluid mechanics research. Indeed, self-similarity is at the heart of theoretical efforts for the modelling of laminar and turbulent boundary layers \citep{Prandtl904,townsend1976structure}, free shear flows \citep{tennekes1972first,townsend1976structure,pope_2000,george1989self}, cascade dynamics \citep{kolmogorov1941c,pope_2000,vassilicos2015dissipation,steiros2022balanced}, linear and nonlinear waves \citep{taylor1950formationA,taylor1950formationB,zel1967physics,whitham2011linear}, singularities \citep{eggers2015singularities}, high Mach number aerofoil design \citep{cantwell2002introduction}, among many others. 

There are three possible strategies to identify self-similarity. First, if the differential equations describing the problem are known and the boundary conditions are simple, one can formally extract the flow symmetries using the theory of Lie groups \citep{cantwell2002introduction,birkhoff2015hydrodynamics, cantwell1978similarity}. Second, if the equations are unknown one may still invoke dimensional analysis to uncover dilational symmetries \citep{barenblatt1996scaling}. Finally, on many occasions, intuition has allowed the uncovering of self-similarity directly, via visual inspection of the problem solution (as for example in the case of turbulent shear flows \citep{townsend1976structure}). However, this is not always sufficient to reveal the underlying self-similarity of a phenomenon. Intuition depends on the skills of the practitioner and is generally confined to simple problems. Dimensional analysis can only treat problems with dimensional variables, while even then it may uncover only the `tip of the iceberg' of possible self-similar solutions, i.e., some (but far from all) of the self-similarities connected to dilational transformations \citep{barenblatt1996scaling}. Lie group theory may become unfeasible in cases where the boundary conditions are overly complicated. Even if this is not the case, it may reveal some but not all symmetries of the problem. A coordinate transformation may be necessary for additional, `hidden' symmetries to be revealed \citep{cantwell2002introduction,liu2022machine}. In other cases the governing equations may be inadequate. An example of the above occurs in turbulent flows, where it is customary to consider the ensemble-averaged flow equations of motion (i.e. the Reynolds-Averaged Navier-Stokes equations). Given the chaotic turbulent dynamics and complex boundary conditions, certain original symmetries of the Navier-Stokes equations are broken, but are still hypothesized to be recovered in a statistical sense \citep{frisch1995turbulence}. However, due to the underdetermined nature of the RANS equations (turbulence closure problem), extraction of symmetries is challenging \citep{oberlack2001unified, oberlack1999similarity, oberlack2006group, oberlack2010new, oberlack2022turbulence}. 
 
The recent emergence of machine learning methodologies has provided a significant boost in our ability to uncover symmetries and self-similarities \citep{desai2022symmetry,yang2023generative,otto2023unified}. If self-similarity exists in a problem, it must appear in the observables (data), and can be, in principle, discovered by data-driven methods. For instance, a known challenge in conventional dimensional analysis (Buckingham Pi theorem) is that it yields a non-unique set of non-dimensional variables that govern the evolution of a physical problem \citep{barenblatt1996scaling,cantwell2002introduction}. Information from data can in that case provide a constraint to Buckingham Pi, and has been recently used to identify the most appropriate set of non-dimensional variables \citep{mendez2004scaling,constantine2017data,jofre2020data,saha2021hierarchical,xie2022data,bakarji2022dimensionally,yuan2025dimensionless}. Other examples are cases in which conservation laws and symmetries of ordinary differential equations cannot be readily extracted because they are hidden, i.e., they require a coordinate transformation before they become manifest. Machine learning can be used to provide the necessary coordinate transformation for hidden symmetries to manifest \citep{liu2022machine,mototake2023extracting}. Of high relevance are also efforts to leverage the information contained in data to reveal scale-invariant flow structures \citep{fukami2024data} or close underdetermined governing equations \citep{duraisamy2019turbulence}, which can support the training and testing of nonlinear machine-learning techniques (e.g., \citet{fukami2024single, duraisamy2025introduction}).
%and potentially open the door for more extended Lie group analysis approaches

In this work, we present a generalized methodology that can identify if similarity variables exist and, if they do, their mathematical expressions directly from data, without prior knowledge of the governing equations or boundary conditions. Our work differs from previous efforts (e.g., \citet{xie2022data,bakarji2022dimensionally,yuan2025dimensionless}) in that it is not based on dimensional analysis (although it is dimensionally consistent) and can thus identify self-similarities beyond dilational transformations connected to dimensional problems. This is achieved by formulating a minimization problem to identify the similarity variables, which are then interpreted analytically via symbolic regression. In particular, our method simultaneously searches for the optimal non-linear transformations of both independent variables (spatiotemporal coordinates) and dependent variables (observables) that yield the similarity variables of the problem and is thus different to methods aiming to identify coordinate transformations that render hidden symmetries manifest (e.g., \citet{liu2022machine}). The paper is structured as follows: in section \ref{sec:methodology} and Appendix \ref{app:algorithm} we describe the algorithm that extracts the similarity variables of a self-similar physical process from data. Section \ref{sec:results} demonstrates the potential utility of the proposed method by applying it to five problems which have been influential in fluid dynamics research and are exactly or approximately self-similar under different transformation types, with data derived from both laboratory and numerical experiments. Finally, Section \ref{sec:discussion} discusses potential applications and limitations of the method and provides concluding remarks. 

\section{Data-driven identification of self-similarity}
\label{sec:methodology}
Let $q(s,t)$ be a quantity of interest governed by a set of non-linear partial differential equations, with $s$ and $t$ being the independent variables, typically associated with a spatial coordinate and time, respectively. In some problems (e.g., a laminar boundary layer), $t$ may also correspond to a spatial coordinate. Consider the set of variables $\xi$ and $\tilde{q}$,  which lead to the transformation $q(s,t) \rightarrow \tilde{q} (\xi,t)$, with $\xi = \xi(s,t)$, where $\xi$ may also depend on the  parameters (constants) of the problem. If $\tilde{q}(\xi,t)$ is independent of $t$, i.e., there is a function $\hat{q}(\xi)$ such that $\tilde{q}(\xi,t) \equiv \hat{q}(\xi)$, then $q(s,t)$ is said to be self-similar \citep{pope_2000}. In this case, the variables and the transformation are referred to as similarity variables and similarity transformation, respectively. Assuming that we have measurements of $q(s,t)$ at distinct instants (stations) $t_i$, where $i=1,\dots,n_t$ and $n_t$ is the number of available instants (stations), we propose a two-stage workflow for extracting the similarity transformation directly from data, without previous knowledge of the governing equations. Because self-similarity can be inexact or the data may include errors, the equivalence $\tilde{q} (\xi,t) \equiv \hat{q}(\xi)$ can be at most expected to hold approximately. In higher dimensions, $s$ and $\xi$ are vectors. 

\subsection*{Step 1. Search for similarity variables}
We express the similarity variables as a superposition of elementary dilation and translation groups, 
\begin{equation}
\xi = \alpha(t) s + \beta(s,t), \quad \tilde{q} = \gamma(t) q + \delta(s,t).
\label{eq:decomp}
\end{equation}
Decomposing the similarity variables in the form of distinct transformations is important for facilitating the success of the ensuing optimization and regression tasks, as well as for enhancing the interpretability of the method. The search for similarity variables is formulated as a minimization problem,
\begin{argmini}|s|
{\boldsymbol{w}}
{\frac{1}{2} \sum_{i = 1}^{n_t} \sum_{j = 1}^{n_t} \| \tilde{q}(\xi,t_i) - \tilde{q}(\xi,t_j) \|_2^2
}
{}{},
\label{eq:optstep1}
\end{argmini}
where $\boldsymbol{w}$ is the design variable matrix containing the discrete values of the transformation functions $\alpha$, $\beta$, $\gamma$, $\delta$. The $l_2$-norms,  $\| \cdot \|_2^2$, are computed following interpolation on the transformed coordinates ($\xi$) grid. The resolution of the $\xi$ grid, i.e., the number of points where the $l_2$ norms are evaluated, can be freely chosen by the user and can  be set equal to the number of points in the input (non-transformed) data (see also Appendix \ref{app:sensitivity}). In some problems, solving Eq. \eqref{eq:optstep1} can lead to degenerate solutions (i.e., $\tilde{q} = 0$), in which case Eq. \eqref{eq:optstep1} is replaced by a mean-regularized cost functional, which prevents trivial optima 
\begin{argmini}|s|
{\boldsymbol{w}}
{\frac{1}{2} \sum_{i = 1}^{n_t} \sum_{j = 1}^{n_t} \bigg{\|} \frac{\tilde{q}(\xi,t_i) - \tilde{q}(\xi,t_j)}{ \sqrt{ | \left( \tilde{q}(\xi,t_i) + \tilde{q}(\xi,t_j) \right)/2| } } \bigg{\|}_2^2
}
{}{}.
\label{eq:optstep1_var}
\end{argmini}
This step provides the discrete values of the functions $\alpha$, $\beta$, $\gamma$, $\delta$, and by extension through Eq. \eqref{eq:decomp}, the discrete values of the similarity variables $\xi$ and $\tilde{q}$, at all $n_t$ instants. In practice, in order to further promote the success of the optimization task whilst also preserving the automated character of the developed method, we propose the implementation of step 1 in successive iterations of increased decomposition complexity (i.e., building gradually from simple dilation ($\beta=\delta=0$) to the  general decomposition ($\xi = \alpha(t) s + \beta(s,t),\; \tilde{q} = \gamma(t) q + \delta(s,t)$)). Depending on the problem (e.g., its boundary conditions), certain candidate transformations might be irrelevant or inadmissible (for example, translation of radial coordinates in an axisymmetric problem). Appendix \ref{app:algorithm} provides the pseudocode describing an example implementation of the method with clarifying remarks.

\subsection*{Step 2. Analytic form of the transformations}
Given the knowledge of $\boldsymbol{w}$, i.e. the values of the functions $\alpha, \beta, \gamma, \delta$ at all instants $n_t$, we employ symbolic regression to extract the analytic form of the transformations. Symbolic regression is a machine learning technique that combines mathematical operators, functions, constants, and state variables to construct a mathematical expression $\psi$ that best represents a given dataset $\mathcal{D}$. In this work, $\mathcal{D}$ refers to the discrete values of each function ($\alpha$, $\beta$, $\gamma$, or $\delta$) that were found in the first step of the workflow. Symbolic regression can identify arbitrary expressions as it does not make any assumptions about the underlying function. The set of variables (library) $\boldsymbol{A}$ which $\psi$ can depend on,  $\psi=\psi(\boldsymbol{A})$, is provided by the user, and is typically composed of the state variables and parameters of the problem.

To reduce the complexity of the regression problem, it is beneficial to take advantage of any prior knowledge regarding the characteristics of the dataset. 
We enforce two properties. The first property that is critical for discovering physical laws is that of dimensional homogeneity: the units of the identified expression $\psi(\boldsymbol{A})$ should match those of the given dataset $\mathcal{D}$. The second property that is relevant to this work is that scale invariance is related to power laws \citep{barenblatt1996scaling}, allowing us in certain cases to restrict the search for expressions in the form of monomials.

This step of the algorithm can be carried out with any symbolic regression method, giving users the flexibility to choose their preferred tool. Here, we demonstrate the workflow with two different symbolic regression methods: the open-source library \texttt{PySR} (\url{https://astroautomata.com/PySR/}), which offers high performance, flexibility, configurability and generality \citep{cranmer2023interpretable}, and a custom, simple regression algorithm for the identification of monomials. In both cases, we use an $l_2$ loss function, with an optional penalization term enforcing dimensional homogeneity
\begin{argmini}|s|
{\psi}
{\| \psi(\boldsymbol{A}) - \mathcal{D} \|_2^2 + w_D \| \left[ \psi(\boldsymbol{A}) \right] - \left[ \mathcal{D} \right] \| } 
{}{},
\label{eq:optstep2}
\end{argmini}
where $w_D$ is a non-negative regularization factor and the brackets $\left[ \; \right]$ denote the units of a quantity in the form of a dimension vector (see Appendix \ref{app:algorithm} for an example). Equation \eqref{eq:optstep2} is individually applied to each transformation function ($\alpha$, $\beta$, $\gamma$, or $\delta$). This step provides the analytic form of the similarity transformations $\alpha$, $\beta$, $\gamma$, $\delta$, and thereby of the similarity variables $\xi$ and $\tilde{q}$ through Eq. \eqref{eq:decomp}.  

\section{Results}
\label{sec:results}

\subsection*{Laminar boundary layer}

\begin{figure}
\centering
\includegraphics[width=\textwidth]{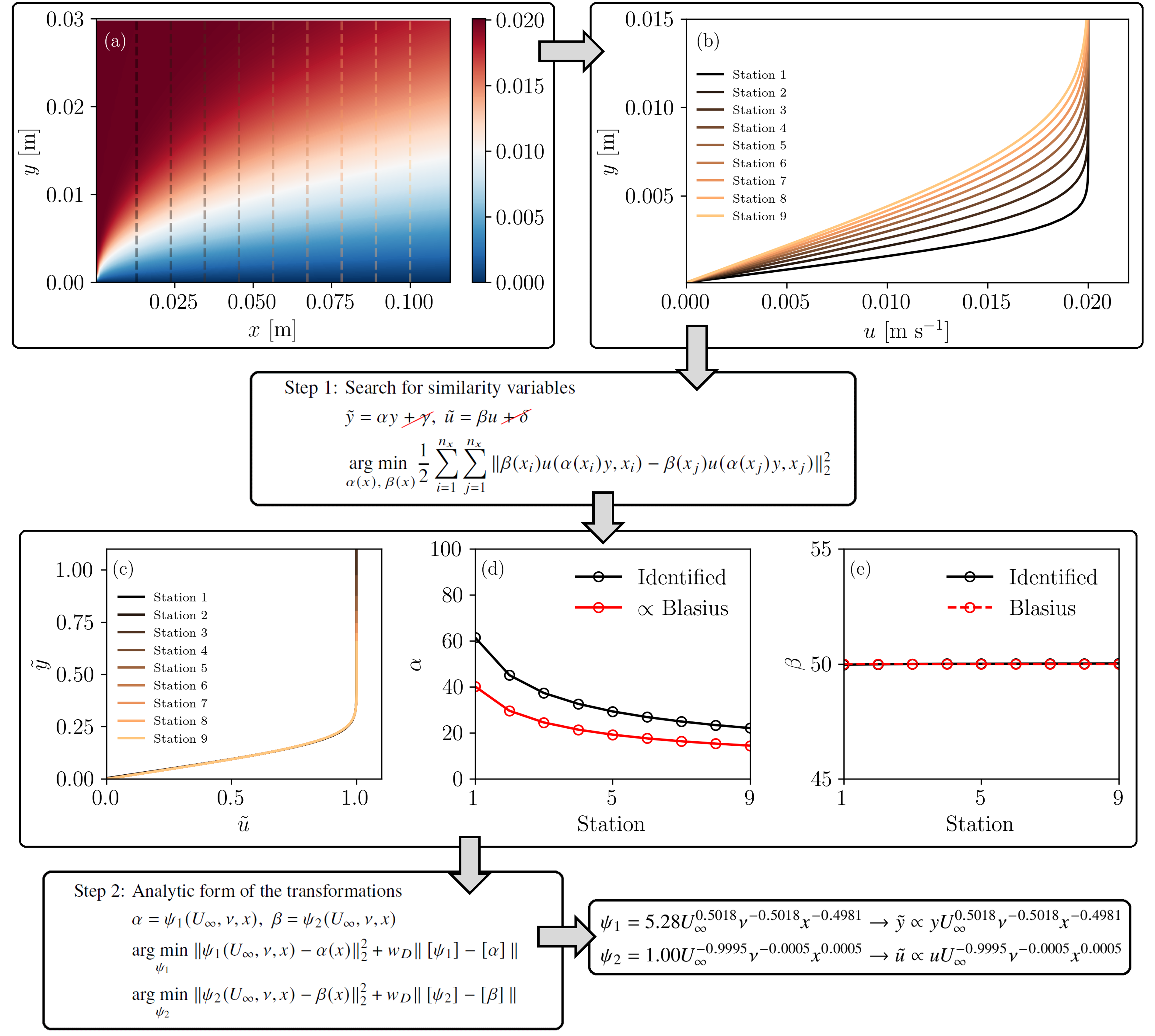}
\caption{Data-driven identification of self-similarity in the Blasius boundary layer. 
(a) Streamwise velocity field (Blasius solution). The dashed vertical lines denote the nine velocity profile sampling stations. 
(b) Sampled velocity profiles. 
(c) Algorithmically collapsed velocity profiles. 
(d) Algorithmically identified scaling $\alpha$ of the wall-normal coordinate $y$. 
(e) Algorithmically identified scaling $\beta$ of the streamwise velocity $u$.
}
\label{fig:bl}
\end{figure}

%The thin layer of fluid in the immediate vicinity of a solid boundary is known as the boundary layer. 
%Knowledge of the boundary layer properties is of enormous importance for many applications in the transportation and energy sectors. 
The first validation example considers the well-known case of the two-dimensional, laminar and incompressible  boundary layer, where \citet{Prandtl904} showed that some terms of the Navier-Stokes equations are negligible, and reduced the latter to the ``Boundary Layer Equations" (BLE).
By utilising the streamfunction $\Psi$, 
%where $\Psi_y = u$ and $\Psi_x = -v$, with $u$ the streamwise flow velocity and $v$ the crosswise velocity, 
the BLE can be expressed in a single equation.
%,  
%\begin{equation}
%\Psi_y \Psi_{xy} - \Psi_x \Psi_{yy} - \nu \Psi_{yyy} =0
%\end{equation}
%where , and the subscripts denote partial differentiation. 
\citet{blasius1908} noted that for semi-infinite flat plate boundary conditions the BLE is symmetrical under a dilational transformation, which reduces the problem to an ordinary differential equation for the nondimensional streamfunction $f(\tilde{y})=\Psi/\sqrt{\nu U_\infty x}$, with $\tilde{y} = y \sqrt{U_\infty/(\nu x)}$. In the above, $U_\infty$ is the free stream velocity away from the flat plate, $x$ and $y$ are the streamwise and normal-to-the-wall distances, respectively, and $\nu$ is the kinematic viscosity of the fluid. The BLE are thus reduced to the boundary value problem
\begin{subequations}
\begin{equation}
2 f_{\tilde{y} \tilde{y} \tilde{y}} + ff_{\tilde{y} \tilde{y}} =0,
\end{equation}
\begin{equation}
f(0) = 0, f_{\tilde{y}}(0) = 0, f_{\tilde{y}}(\infty) = 1.
\end{equation}
\end{subequations}
Knowledge of $\tilde{u} = f_{\tilde{y}}$ (e.g., via numerical integration of the Blasius equation) allows for the calculation of the streamwise velocity distribution $u = U_\infty \tilde{u}$ at any location in the boundary layer. Here, we derive the Blasius similarity directly from data, without prior knowledge of the Navier-Stokes equations, Prandtl's scaling analysis, or Blasius' similarity arguments. 
Our dataset consists of $u(x,y)$ data from a flat plate laminar boundary layer flow (figure \ref{fig:bl}(a)), acquired by solving the Blasius ODE with a shooting method \citep{bakarji2022dimensionally}. In particular, we extract velocity profiles $u(y)$ from nine stations at different streamwise coordinates $x$ (figure \ref{fig:bl}(b)). The algorithm commences by seeking similarity under the dilational transformations $\tilde{y} = \alpha(x) y$, $\tilde{u} = \beta(x) u$. Figure \ref{fig:bl}(c) shows that the transformed profiles have collapsed onto a single curve. The identified transformations for the wall-normal coordinate $y$ and the streamwise velocity $u$ are shown in figures \ref{fig:bl}(d) and (e), respectively. 

The second step consists of expressing the identified transformations as functions of the governing parameters and independent variables, i.e., a symbolic regression task. The library of candidate variables is composed of the free-stream velocity $U_\infty$, the viscosity $\nu$, and the streamwise coordinate of the extracted profiles $x$, i.e., $\alpha= \alpha(U_\infty,\nu,x)$ and $\beta=\beta(U_\infty,\nu,x)$. Owing to the nature of the assumed dilational transformations, we look for expressions in the form of monomials, i.e., $\alpha=c_1 U_\infty^{c_2} \nu^{c_3} x^{c_4}$ and $\beta=c_5 U_\infty^{c_6} \nu^{c_7} x^{c_8}$, where $c_1$ and $c_5$ are multiplicative constants that reflect the arbitrariness of the dilation factors between the original and similarity variables. Dimensional homogeneity is enforced by assuming $\tilde{y}$ and $\tilde{u}$ to be dimensionless and a non-zero weighting constant $w_D$ in the objective function (Eq. \eqref{eq:optstep2}). The interpreted expressions are
\begin{subequations}
\begin{equation}
\tilde{y} \propto y U_\infty^{0.5018} \nu^{-0.5018} x^{-0.4981}
\end{equation} 
\begin{equation}
\tilde{u} \propto u U_\infty^{-0.9995} \nu^{-0.0005} x^{0.0005}
\end{equation} 
\end{subequations}
which are in close agreement with the theoretically derived similarity variables of Blasius ($\tilde{y} = y \sqrt{U_\infty/(\nu x)}$, $\tilde{u} = u / U_\infty$). 
As previously noted, the identified and theoretical scalings can vary by an arbitrary multiplicative constant (i.e., the offset in figure \ref{fig:bl}(d)), depending as to whether and how the user normalizes the collapsed profiles. Appendix \ref{app:sensitivity} shows results from a sensitivity and robustness analysis with respect to the input data, parameters, and noise.   
  
\subsection*{Burgers' equation}

\begin{figure}
\centering
\includegraphics[width=\textwidth]{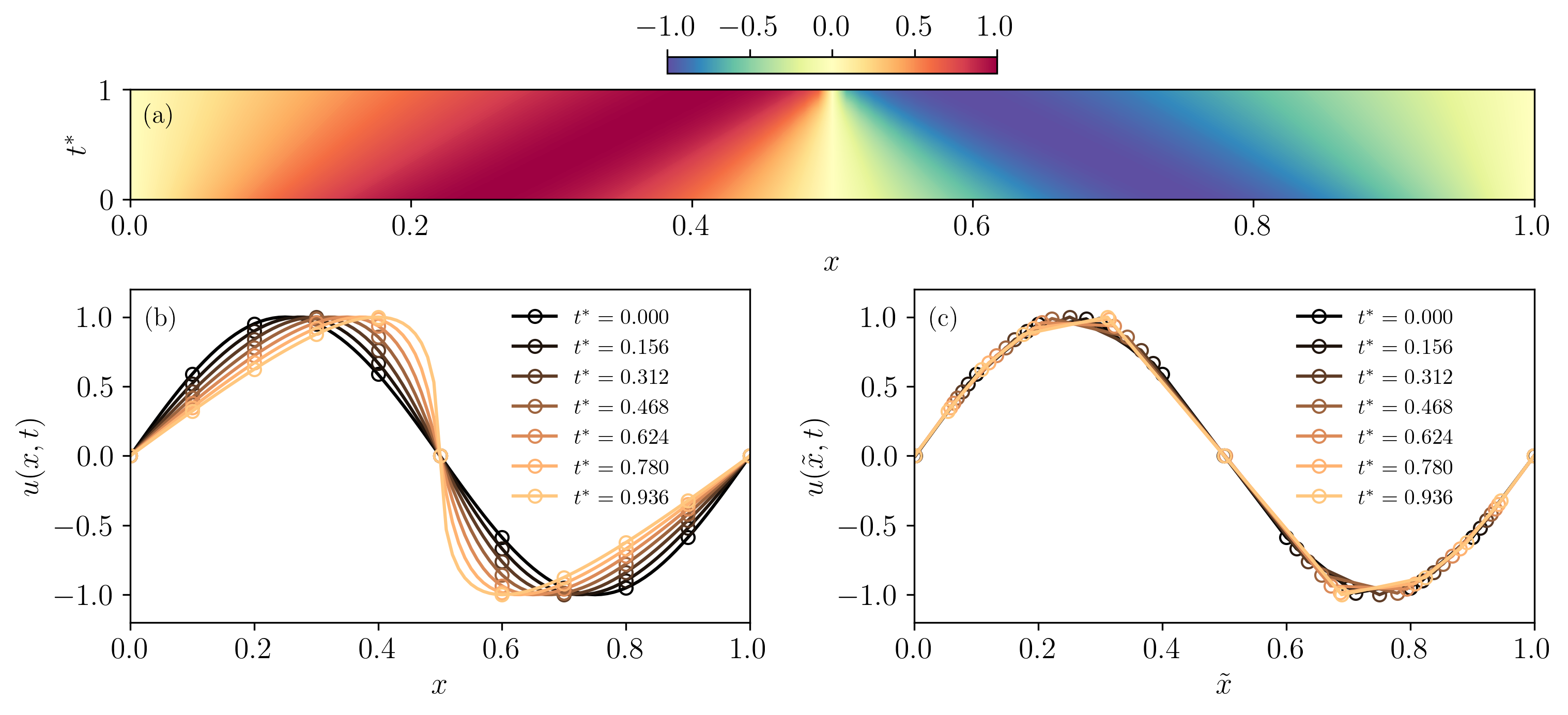}
\caption{Data-driven identification of self-similarity in Burgers' equation. 
(a) Spatiotemporal evolution of the velocity. 
(b) Extracted profiles at different time instants. Markers show the data given to the algorithm. 
(c) Algorithmically collapsed profiles.}
\label{fig:burgers}
\end{figure}

The second validation example considers Burgers' equation, which is a partial differential equation that finds wide application in fluid dynamics, non-linear acoustics, and traffic flow, among others. It acts as a prototype for a variety of phenomena including shock wave formation, rarefaction waves, and turbulence \citep{zel1967physics,whitham2011linear}. 
In the absence of diffusion, it is the simplest model for gasdynamics.  
Its mathematical expression in one dimension, along with its initial condition is 
\begin{equation} 
u_t + u u_x = 0, \quad u(x,0) = f(x)
\label{eq:burgers}
\end{equation}%
where $u(x,t)$ is the fluid velocity and $x$ and $t$ the spatial and temporal coordinates, respectively. The subscript denotes partial differentiation. Using the method of characteristics, the solution can be expressed as $u(x,t) = f(\xi) = f(x-ut) = f(x+\alpha)$, i.e., the problem described by Eq. \eqref{eq:burgers} is invariant under a non-uniform translational transformation. In this work, we consider a sinusoidal initial condition, $u(x,0) = \sin{(2 \pi x)}$, with $x \in \left[0,1\right]$. Figures \ref{fig:burgers}(a) and (b) show the spatiotemporal evolution of velocity and profiles extracted at seven different time instants, respectively, illustrating the steepening of the solution as time evolves. The times shown are non-dimensionalised with the time of shock formation $t_c$, $t^*=t/t_c=2 \pi t$. The algorithm is provided with the data shown with markers in figure \ref{fig:burgers}(b).

Figure \ref{fig:burgers}(c) shows the algorithmically collapsed velocity profiles on the transformed coordinates $\left(\tilde{x},t \right)$, i.e. the output of Step 1 of the algorithm, following a transformation of the form $\tilde{x} = x+ \alpha(x,t)$ (simpler transformations do not succeed at collapsing the data). Since this is an initial value problem, we only consider the distance of the profiles with respect to the initial condition, i.e., the objective function in Eq. \eqref{eq:optstep1} is simplified to $\sum_{i=1,\dots,n_t} \| u(\tilde{x},t_i) - u(x,0) \|_2^2$. 

We use {\texttt{PySR}} to interpret the algorithmically computed transformation $\alpha$. The library of candidate variables is composed of the problem's independent and dependent variables, i.e., $\alpha= \alpha(u,x,t)$, and the library of operators includes the four basic mathematical operations ($+,-,\times,\div$). $\texttt{PySR}$ interprets the algorithmically computed transformation as $\alpha = - u t$, which is identical to the analytical similarity transformation. Close inspection of figure \ref{fig:burgers}(c) shows that the profiles are not perfectly collapsed (for example, at $t^*=0.936$). The range-normalised mean absolute error between the algorithmically computed discrete values of the transformation and the analytical ones is $2.2\%$. However, in Step 2 of the workflow (symbolic regression), the algorithm attempts to fit expressions that balance complexity and accuracy, allowing it to recover the exact analytical transformation $\alpha = -ut$.  

\subsection*{Free turbulent flow}
\begin{figure}
\centering
\includegraphics[width=\textwidth]{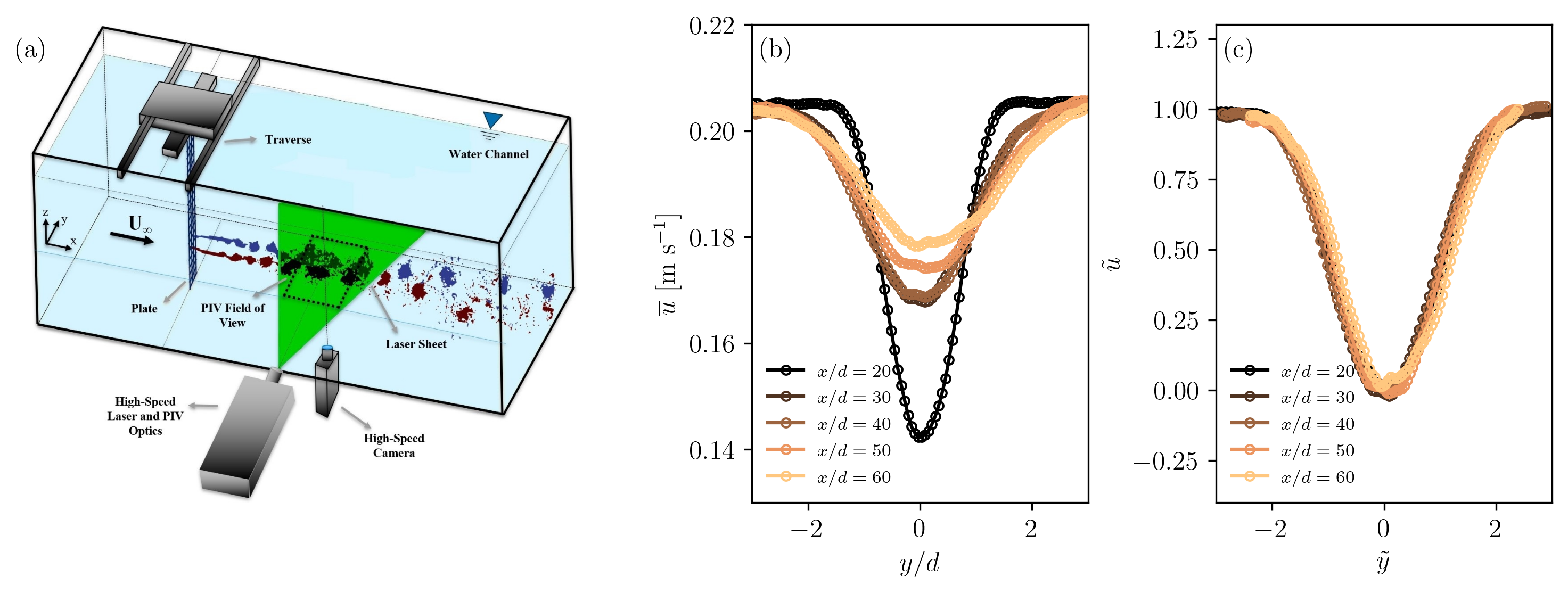}
\caption{Data-driven identification of self-similarity in the wake of a porous plate. 
(a) Schematic of the experimental apparatus showing the flume, porous plate and PIV configuration. 
(b) Mean streamwise velocity profiles at different locations downstream of the plate. 
(c) Algorithmically collapsed velocity profiles.}
\label{fig:wake}
\end{figure}

As a third example, we consider the statistically stationary flow past a slender bluff body at high Reynolds numbers, which is a characteristic example of a free (i.e., unconfined) turbulent shear flow. The term slender signifies an effectively infinite body length in one crossflow direction, along which the flow statistics can be considered homogeneous. 
Such flows can be modelled using a boundary layer approximation, similar to the one employed by Prandtl for laminar boundary layers, applied to the time-averaged flow \citep{townsend1976structure}, i.e., 
\begin{subequations}
\label{eq:rans}
\begin{equation}
\bar{u}_x + \bar{v}_y = 0
\end{equation}
\begin{equation}
\bar{u} \bar{u}_x + \bar{v} \bar{u}_y = \nu \bar{u}_{yy} - (\overline{u'v'})_y
\end{equation}
\end{subequations}
where the overbar denotes time-averaging and $\nu$ is the kinematic viscosity of the fluid. $u$ and $v$ are the flow velocities along the streamwise $x$ and crosswise $y$ directions, which are decomposed into time-averaged $(\bar{u},\bar{v})$ and fluctuating $(u',v')$ components. The formulation leading to Eq. \eqref{eq:rans} introduces the Reynolds stress term $\overline{u'v'}$ into the flow governing equations, i.e., an additional unknown which cannot be calculated implicitly (this is the well-known closure problem of turbulence). The identification of similarity in this case is, therefore, not possible from the equations themselves, unless an assumption is made regarding the relation of the unknown Reynolds stress terms to the mean velocity distribution (i.e., turbulence modelling). However, it is customary to hypothesize that the turbulent boundary layer equations accept self-similar solutions far from initial conditions, a fact also supported by experimental observations \citep{townsend1976structure,cantwell2002introduction}. This approximate self-similarity is the origin behind various scaling laws for the evolution of free shear flows, widely used in many applications of the energy and transportation sectors, such as wind farm planning \citep{bempedelis2022analytical} and jet engine noise prediction \citep{tam2019phenomenological}.
Returning to the particular example of the slender bluff body wake, we may consider the time-averaged velocity deficit $\zeta(x,y) = U_\infty - \bar{u}(x,y)$ that the body produces far (i.e., tens of characteristic body lengths) downstream. In the above, $U_\infty$ is the constant free stream velocity. The flow statistics are homogeneous along the spanwise direction $z$. Wake self-similarity is then known to assume the form \citep{pope_2000,townsend1976structure}
\begin{equation} 
\zeta(x,y) = \tilde{\zeta}(\xi) (U_\infty -\bar{u}_{\text{cntr}}(x)), \quad \text{with } \xi= \frac{y}{y_{1/2}(x)}\,,
\label{eq:self-wake}
\end{equation}
where $\bar{u}_{\text{cntr}}=\bar{u}(x,0)$ denotes the centreline velocity and $y_{1/2}(x)$ is the wake half-width defined such that $\bar{u}(x,\pm y_{1/2}) = 0.5 \left(U_\infty +u_{\text{cntr}}\right)$.
We attempt to extract the self-similar relation Eq. \eqref{eq:self-wake} directly from experimental data. To this end, we measured the turbulent wake of a plate of 53\% porosity immersed in a water flume normal to the flow (see figure \ref{fig:wake}(a)) at a Reynolds number based on the free stream velocity and plate width $Re \approx 6,000$. The velocity fields at various positions downstream of the plate were measured using Particle Image Velocimetry (Phantom 4MP camera at 50 Hz acquisition frequency) as shown in figure \ref{fig:wake}(a). Each mean streamwise velocity profile was then calculated by averaging 3,000 vector fields. More information regarding the experimental procedure can be found in \citet{ElifReview, bekogluSVS}. 
We consider the mean velocities at five stations in the wake of the plate (figure \ref{fig:wake}(b)). Figure \ref{fig:wake}(c) shows the collapse that is obtained via the proposed method, assuming transformations of the form $\tilde{y} = \alpha(x) y$ and $\tilde{u} = \beta(x) \bar{u} + \gamma(x)$ (simpler transformations cannot collapse the data). The identified transformations are regressed as functions of the characteristic scales and variables of the problem using $\texttt{PySR}$, with $\alpha = \alpha(x, y_{w}, y_{1/2})$ (with $y_w(x)$ defined as $\bar{u}(x,\pm y_{w}) = 0.99 U_\infty$), $\beta=\beta(U_\infty, \bar{u}_{\text{cntr}}, x, \nu)$ and $\gamma=\gamma(U_\infty, \bar{u}_{\text{cntr}}, x, \nu)$. The library of operators consists of the four basic mathematical operations ($+,-,\times,\div$). The interpreted similarity transformations are 
\begin{subequations}
\begin{equation}
\tilde{y}= \alpha y = \frac{1.0605}{y_{1/2}(x)} y
\end{equation}
\begin{equation}
\tilde{u}(x,\tilde{y}) = \beta\bar{u} + \gamma = \frac{1}{U_\infty - \bar{u}_\text{cntr}} \bar{u} - \frac{\bar{u}_\text{cntr}}{U_\infty - \bar{u}_\text{cntr}} = 1 - \tilde{\zeta}
\end{equation}
\end{subequations}
which match (ignoring the arbitrary multiplicative constant) the self-similar expressions for turbulent wakes found in the literature \citep{tennekes1972first,pope_2000,townsend1976structure}. Besides our own experiments, we also test our method by using experimental data available in the literature \citep{cimbala1988large}, and again retrieve the self-similarity relations for turbulent wakes (see Appendix \ref{app:wake}). 

\subsection*{Cavity collapse}

\begin{figure}
\centering
\includegraphics[width=\textwidth]{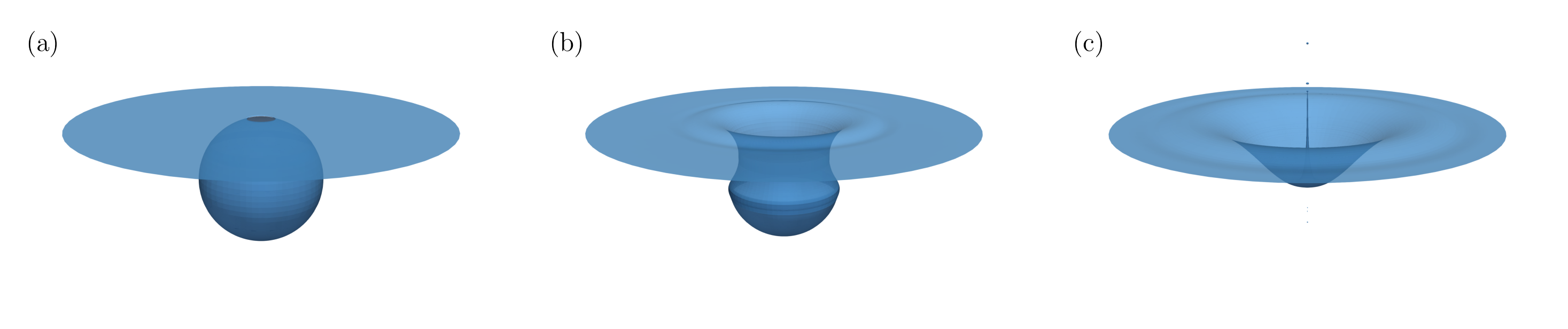}
\includegraphics[width=\textwidth]{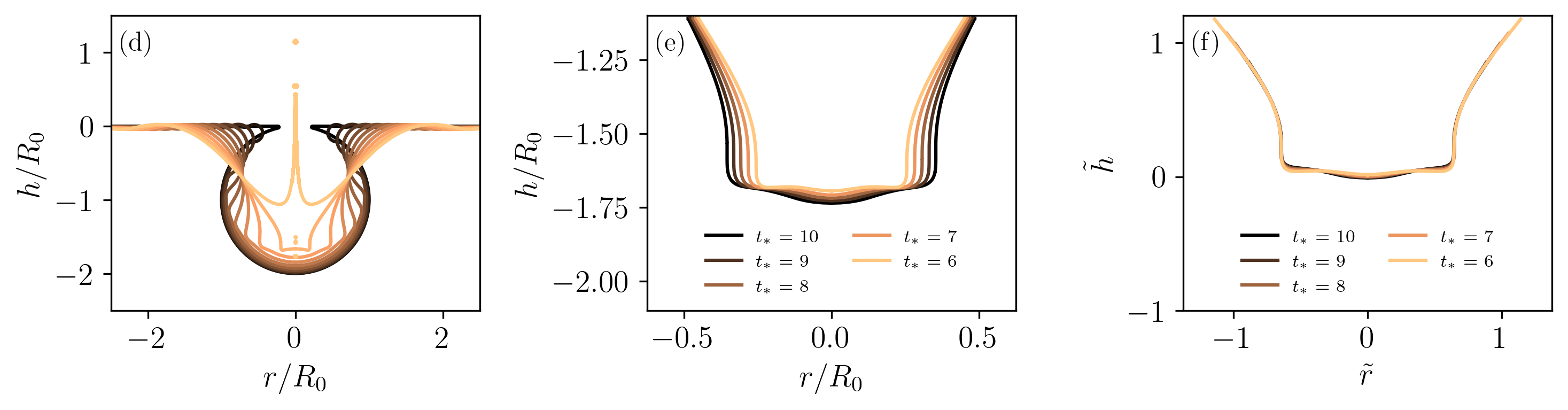}
\includegraphics[width=\textwidth]{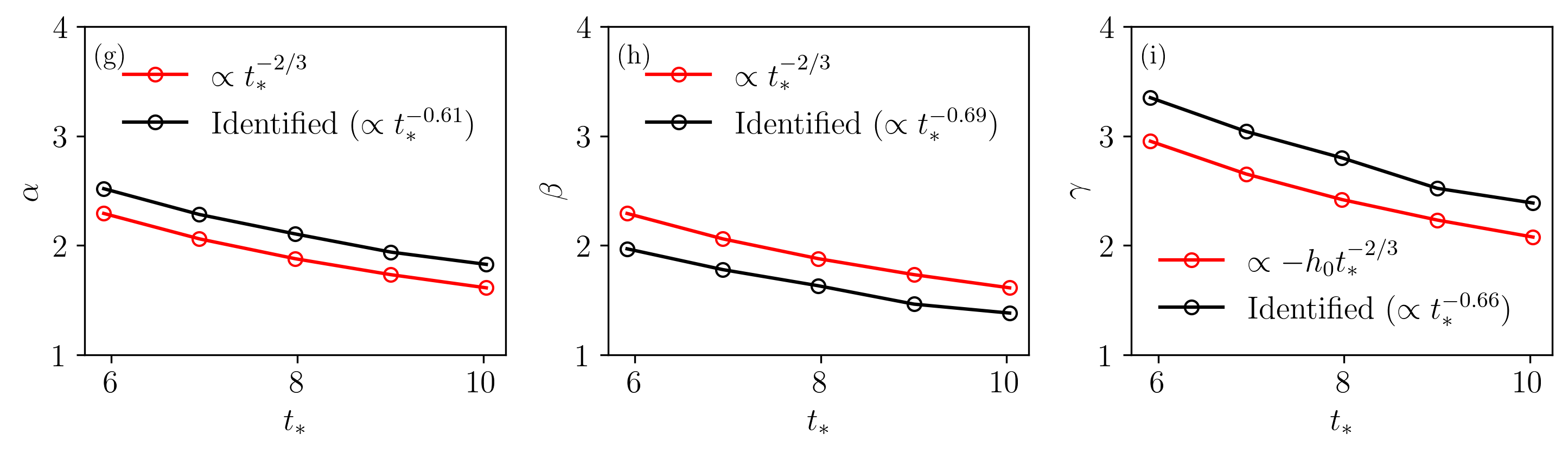}
\caption{Data-driven identification of self-similarity in a collapsing cavity. Three-dimensional visualization of the liquid-gas interface at (a) $t/\tau=0$, (b) $t/\tau=0.25$, and (c) $t/\tau=0.5$, where $\tau=\sqrt{\rho R_0^3 / \gamma}$ is the inertio-capillary timescale. (d) Liquid-gas interface time evolution, $t/\tau = \left(0, 0.05, \dots, 0.5 \right)$. (e) Interface profiles near cavity collapse. (f) Algorithmically collapsed interface profiles near cavity collapse. (g-i) Identified transformations $\alpha$, $\beta$, and $\gamma$. Comparison with theoretical scaling laws.}
\label{fig:bubble}
\end{figure}

The bursting of bubbles at the sea surface plays a crucial role in the exchanges between oceans and the atmosphere, thereby being  important for  climate and weather \citep{deike2022mass}. When a bubble bursts at the sea surface, aerosol is produced via two mechanisms: the rupture of the bubble's cap film, and the formation of a vertical jet that breaks into droplets following cavity collapse \citep{deike2022mass}. The jet and aerosol properties are a consequence of the non-linear fluid dynamics near the points where topology changes \citep{eggers1997nonlinear,eggers2015singularities}. In the case of a collapsing cavity, several studies have found that near the collapse of the travelling capillary waves at the axis of symmetry, the liquid-gas interface evolves in a self-similar manner, enabling the derivation of scaling laws for the bubble and jet dynamics \citep{keller1983surface,zeff2000singularity,duchemin2002jet,ghabache2014physics,ganan2017revision,deike2018dynamics}. 

We simulate the collapse of a cavity ($\text{Bo}=\rho g R_0^2/\gamma = 10^{-3}$, ${\text{La}}=\rho \gamma R_0/ \mu^2 = 2500$) by solving the two-phase incompressible axisymmetric Navier-Stokes equations using Basilisk (\url{www.basilisk.fr}), an open-source flow solver that has been previously used in computational studies of bursting bubbles \citep{lai2018bubble,berny2020role,sanjay2021bursting}. In the above, $\rho$ and $\mu$ are the liquid density and viscosity, respectively, $\gamma$ is the interfacial tension, $R_0$ is the initial bubble radius, and $g$ is the gravitational acceleration. The evolution of the liquid-gas interface is shown in figures \ref{fig:bubble}(a-d). We extract profiles of the interface $h(r,t)$ as it approaches the axis of symmetry, at $t_* = \left(t_0 - t \right)/t_c = \left[10, 9, 8, 7, 6 \right]$, where 
%$t_c = \phi^{3/2} \ell_\mu / V_\mu$ 
$t_c$ is the characteristic time of the horizontal capillary wave \citep{lai2018bubble}, and $t_0$ is the moment where the capillary waves meet at the axis of symmetry, $r=0$. The extracted profiles are shown in figure \ref{fig:bubble}(e). \citet{duchemin2002jet,lai2018bubble} observed that in this time window, the interface approximately follows the $\left(t_0-t\right)^{2/3}$ scaling of inviscid theory \citep{keller1983surface}.
The algorithm successfully collapses the extracted profiles into a single curve (figure \ref{fig:bubble}(f)) for transformations of the form $\tilde{r} = \alpha(t) r$ for the radial coordinate and $\tilde{h} = \beta(t) h + \gamma(t)$ for the liquid-gas interface. By regressing the identified transformations $\alpha$, $\beta$, and $\gamma$ as power-law functions of the non-dimensional time $t_*$ (figure \ref{fig:bubble}(g-i)), we retrieve the expressions
\begin{subequations}
\begin{equation}
\tilde{r} = c_1r t_*^{-0.61}
\end{equation}
\begin{equation}
\tilde{h} = c_2 h  t_*^{-0.69} - c_3 h_0 t_*^{-0.66}
\end{equation}
\end{subequations}
which are close (ignoring the arbitrary multiplicative constants $c_1$, $c_2$, and $c_3$) to the theoretically derived ones \citep{keller1983surface,zeff2000singularity,duchemin2002jet,ghabache2014physics,lai2018bubble}. 

\subsection*{Decaying turbulence}
\begin{figure}
\centering
\includegraphics[width=.66\textwidth]{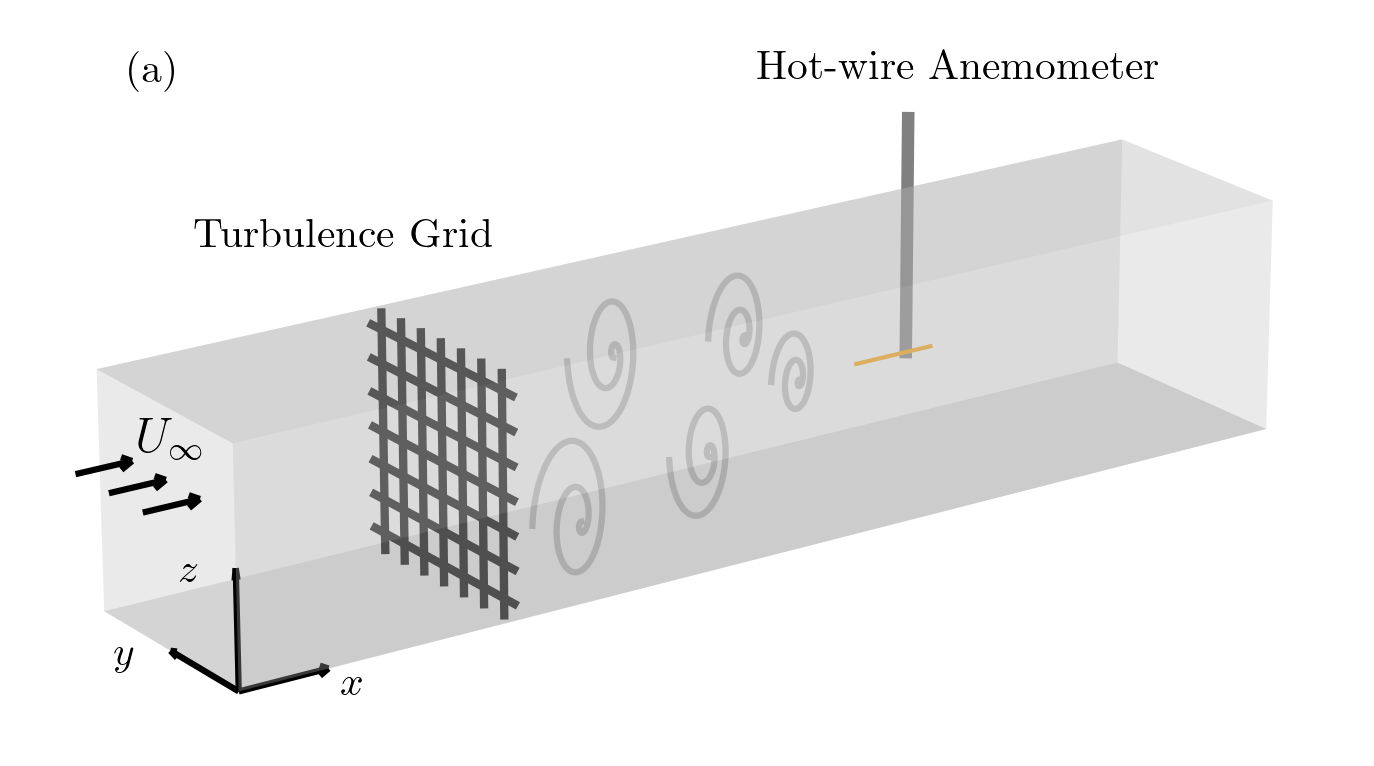}
\includegraphics[width=0.48\textwidth]{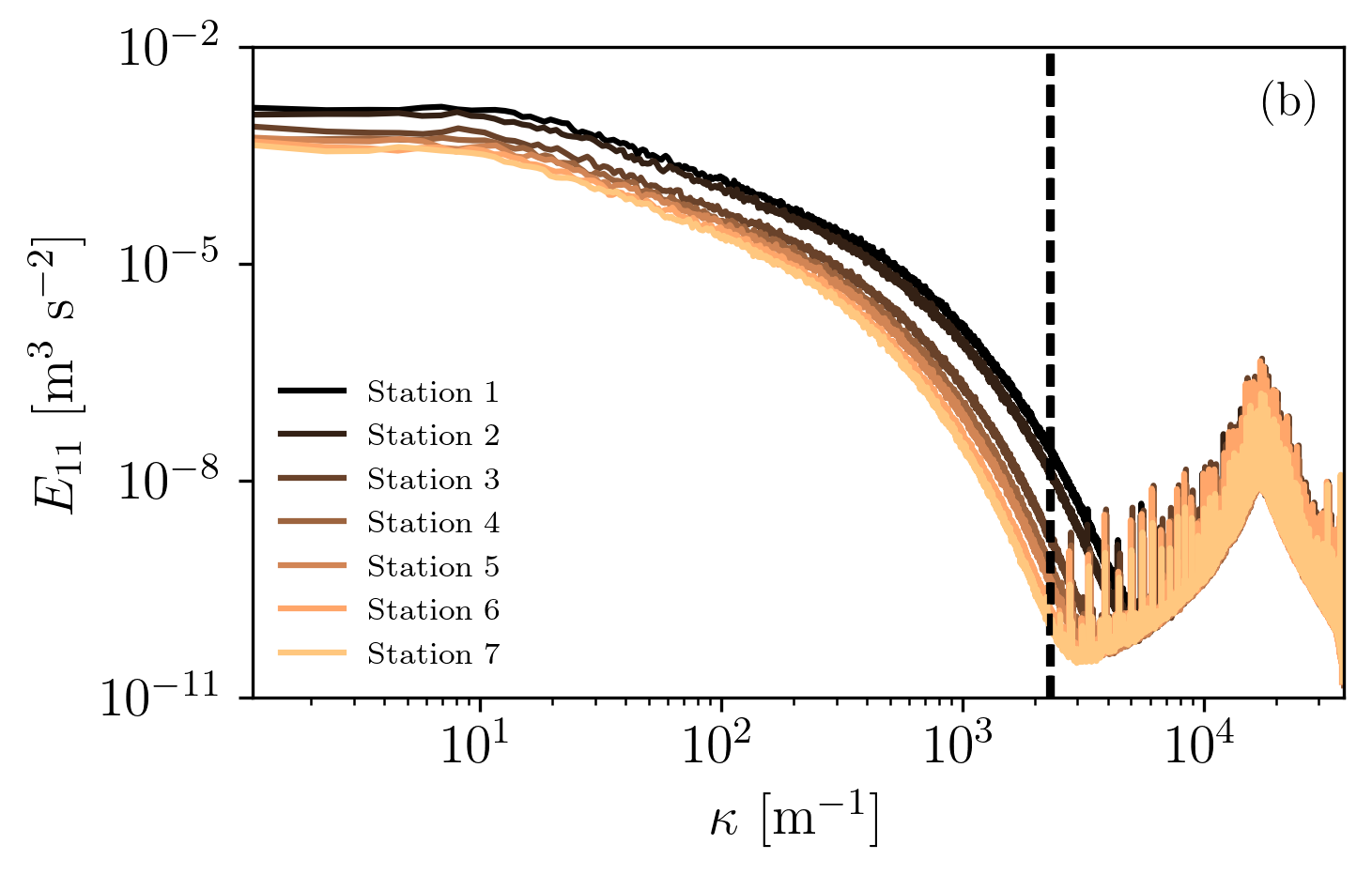}
\includegraphics[width=0.48\textwidth]{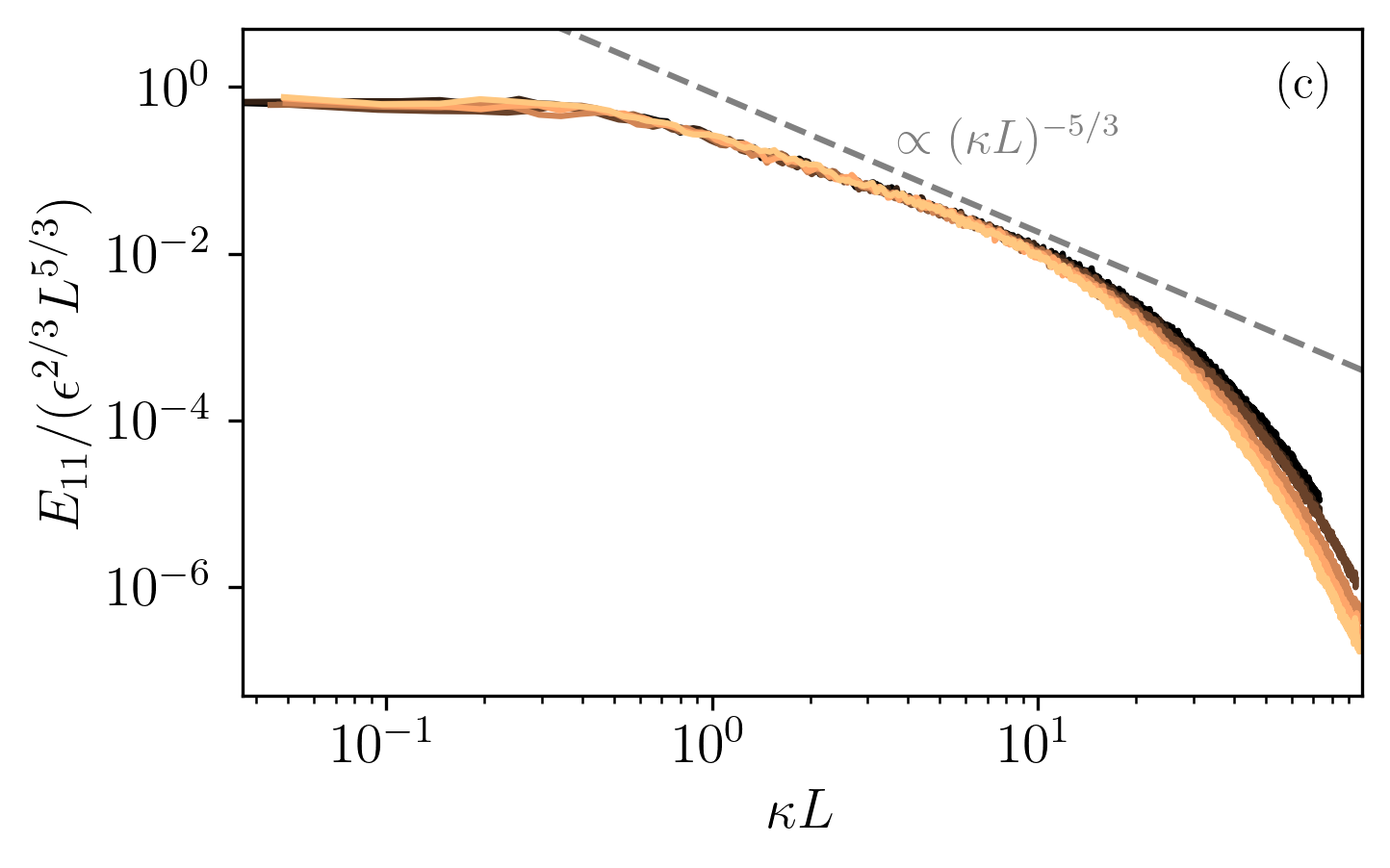}
\includegraphics[width=0.48\textwidth]{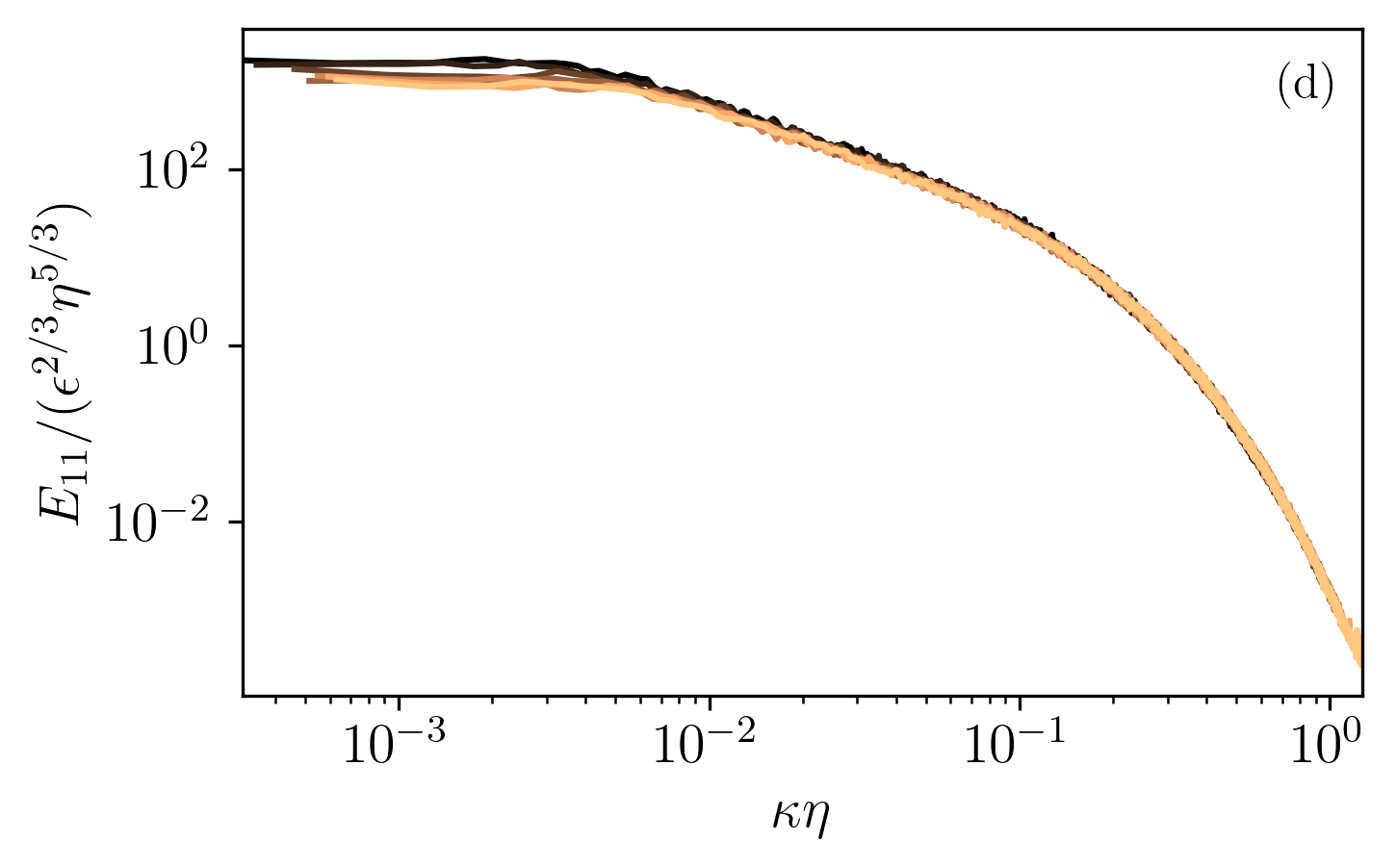}
\includegraphics[width=0.48\textwidth]{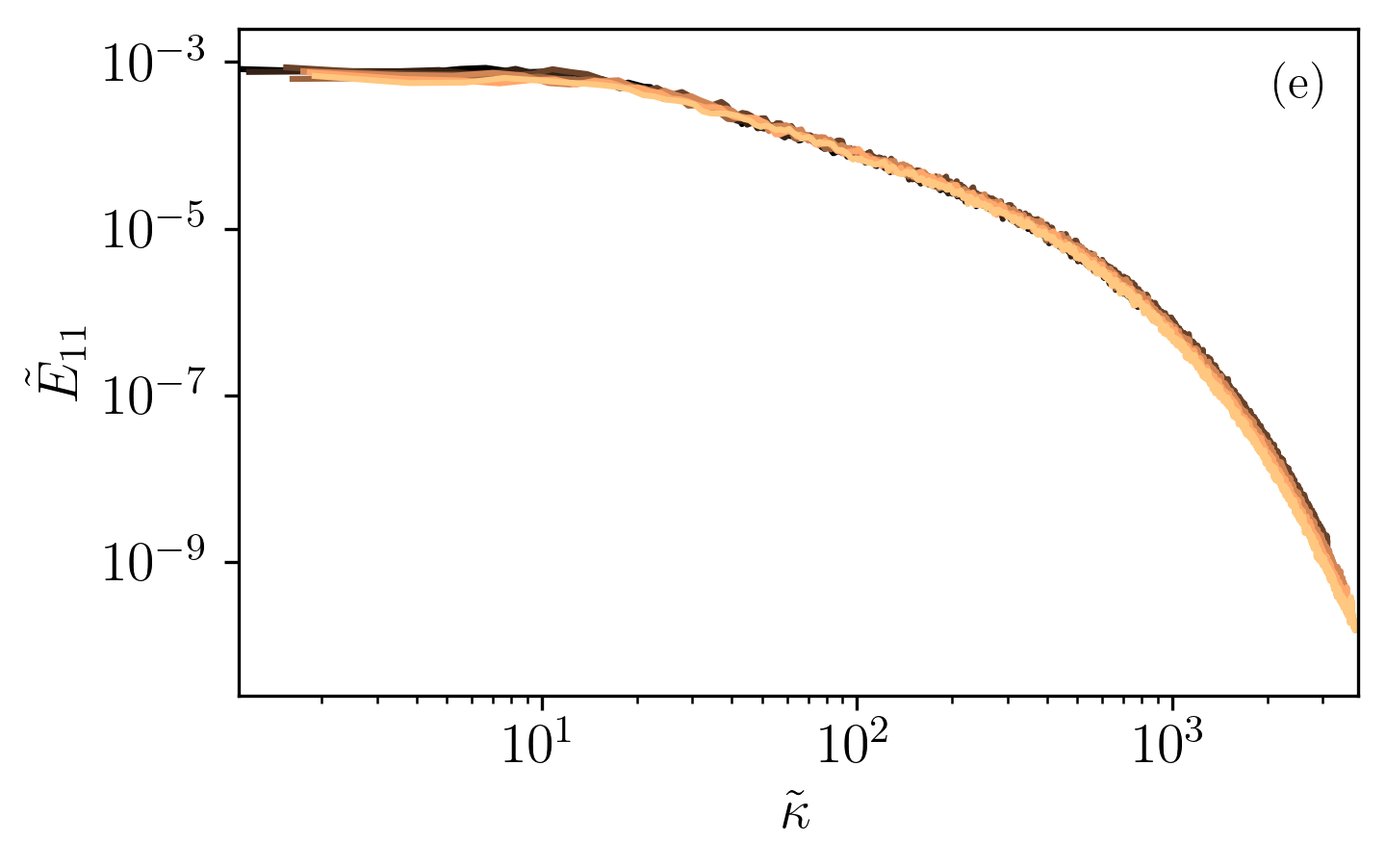}
\caption{Data-driven identification of self-similarity in decaying turbulence. (a) Schematic of the experimental set-up. (b) Experimentally measured power spectral densities. The dashed vertical line delineates the range of the spectrum that is used. (c) Measured spectrum normalized by the inertial scales. (d) Measured spectrum normalized by the Kolmogorov scales. (e) Measured spectrum normalized by the algorithmically identified expression (Eq. \eqref{eq:compositeturb}).}
\label{fig:grid_turbulence}
\end{figure}

We consider the case of homogeneous decaying turbulence, experimentally realized by passing a stream of fluid through a uniformly spaced grid inside a wind tunnel (figure \ref{fig:grid_turbulence}(a)). We analyze measurements of velocity time-series, obtained via hot-wire anemometry at seven positions downstream of the grid (the details of the experimental campaign can be found in \citet{steiros2022balanced}). The measured turbulence is fully developed, approximately homogeneous and yields the -5/3 law for the energy spectrum at intermediately-sized eddies, as predicted by Kolmogorov's K41 framework \citep{kolmogorov1941a,kolmogorov1941b,kolmogorov1941c}.

The theoretical derivation of the -5/3 law assumes that at sufficiently high Reynolds numbers, an intermediate self-similar region forms in the cascade, which is independent of large- and small scale effects \citep{pope_2000,batchelor1953theory,frisch1995turbulence}. However, the above description is part of a more complex flow-picture, as the widely studied paradigm of grid turbulence sufficiently far from initial conditions shows. The turbulence cascade in that case is generally accepted to be characterized by two self-similarities, both present at the same time and at different eddy sizes: One at large scales where viscosity is negligible \citep{steiros2022balanced,lundgren2003kolmogorov}, and one at small scales where non-equilibrium effects are negligible \citep{lundgren2003kolmogorov,pope_2000}. The energy spectrum thus accepts the following general expression \citep{pope_2000}
\begin{equation}
E_{11}(\kappa,x) = \epsilon(x)^{2/3} \kappa^{-5/3} f(\kappa L) g(\kappa \eta)\,,
\label{eq:general}
\end{equation}
where $x$, $\kappa$, $E_{11}(\kappa,x)$, $\epsilon(x)$ represent the distance from the grid, wavenumber, energy spectrum and dissipation rate, respectively. $L$ and $\eta$ are the integral and Kolmogorov scales, characteristic of the large- and small-scale self-similarities, respectively. To retrieve the -5/3 law, $E_{11}(k,x) = \epsilon(x)^{2/3} \kappa^{-5/3}$, one needs to be asymptotically far from large scales (i.e., $\kappa L \to \infty$) and far from small scales ($\kappa \eta \to 0$) at the same time, as in that case K41 assures that both $f$ and $g$ tend to unity. It is of interest to see how our algorithm fares in this two-scale problem.

Figure \ref{fig:grid_turbulence}(b) plots the measured power spectral densities versus the wavenumbers at different measurement stations. Figures \ref{fig:grid_turbulence}(c,d) show the collapse of the power spectral densities using large and small scale similarity variables, respectively. Using a library that consists of $(L, \eta)$ for the $x$ axis and $(\epsilon, L, \eta, k)$ for the $y$ axis, symbolic regression of the algorithmically identified dilational transformations ($\tilde{\kappa} = \alpha(x) \kappa$, $\tilde{E_{11}} = \beta(x) E_{11}$) yields the following expressions, plotted in figure \ref{fig:grid_turbulence}(e)
\begin{equation}
\label{eq:compositeturb}
\tilde{\kappa} \propto \kappa L^{0.368} \eta^{0.632}, \qquad \tilde{E_{11}} \propto E_{11} \epsilon^{0.695} L^{0.709} \eta^{-1.015} k^{-2.042}.
\end{equation}
As expected, the algorithm cannot derive the two-scale similarity expression \eqref{eq:general}, as the assumed dilational transformations effectively reduce it to a single-scale algorithm. Further insight may be obtained by using the dissipation scaling $\epsilon \propto k^{3/2}/L$, where $k$ is the turbulence kinetic energy, which is known to characterize homogeneous decaying turbulence far from initial conditions \citep{steiros2022balanced,steiros2022turbulence,vassilicos2015dissipation}. Expression \eqref{eq:compositeturb} then becomes 
\begin{equation}
\label{eq:compositeturb2}
\tilde{\kappa} \propto \kappa L^{0.368} \eta^{0.632}, \qquad \tilde{E_{11}} \propto E_{11} \epsilon^{-0.667} \left( L^{0.391} \eta^{0.609} \right)^{-5/3}.
\end{equation}
Inspection of the above expression reveals that our algorithm has, seemingly, collapsed the totality of cascade dynamics based on a single empirical length scale, $l \approx L^{0.38} \eta^{0.62}$, which is a combination of $L$ and $\eta$. There have been several attempts to develop single length scale theories of turbulence, derived by assuming self-similarity of the turbulence statistics, i.e., see \cite{de1938statistical}, \cite{sedov2018similarity}, and the later  theories of \cite{barenblatt1974theory} and  \cite{george1992decay}. The last two studies predict the Taylor microscale as the appropriate similarity variable, in close agreement with our data-driven methodology. In homogeneous decaying turbulence the Taylor microscale becomes $\lambda = L^{1/3} \eta ^{2/3}$ \citep{pope_2000}, which is  close to our empirically-derived length scale.

It might be tempting to interpret our empirical results as being in support of a single-scale picture for homogeneous turbulence, contrary to the commonly accepted view of turbulence as an intrinsically multi-length scale phenomenon \citep{batchelor1953theory,townsend1976structure,tennekes1972first,pope_2000}. However, the results shown in figure \ref{fig:grid_turbulence} (and in most studies of homogeneous decaying turbulence) consider the far region of a turbulence grid where the integral length Reynolds number evolution is relatively slow, leading to a similarly slow evolution of $L/\eta$ \citep{vassilicos2015dissipation}. As a result, our data shows that all length scales in question (integral, Kolmogorov, Taylor) produce a (visually) `good-enough' collapse, as shown in figures \ref{fig:grid_turbulence}(c)-(e), and inspection of any of these graphs might give the impression of a single-length scale self-similar evolution.

\begin{figure}
\centering
\includegraphics[width=\textwidth]{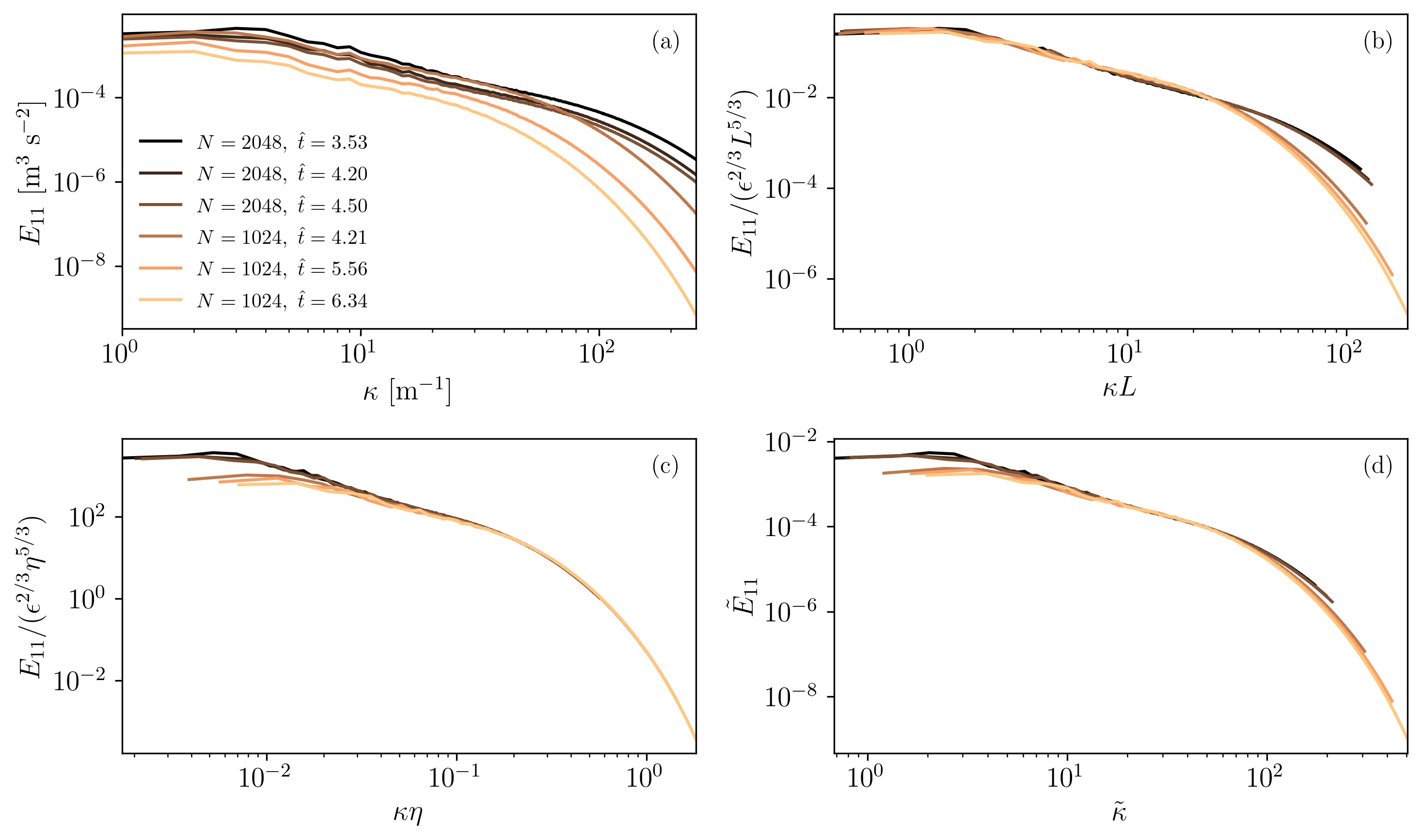}
\caption{Data-driven identification of self-similarity in decaying turbulence, using the DNS data of \citet{goto2016unsteady}. (a) Power spectral densities. (b) Spectrum normalized by the inertial scales. (c) Spectrum normalized by the Kolmogorov scales. (d) Spectrum normalized by the algorithmically identified expression (Eq. \eqref{eq:dnsgoto}).}
\label{fig:DNS_spectra}
\end{figure}

To better appreciate the single- or multi-length scale nature of the turbulence cascade, we need data for which the integral length scale Reynolds number varies drastically. 
To achieve this, we analyse the periodic box DNS data of decaying turbulence taken from \cite{goto2016unsteady} (the details of the simulations can be found in the reference). Similar to the grid case, our goal is to collapse energy spectra taken at different decay times $\hat{t}$ (see figure \ref{fig:DNS_spectra}(a)). Here, however, we consider datasets from two simulations which are identical in every aspect, apart from their size ($N=1024^3$ and $N=2048^3$) and the kinematic viscosity of the fluid, i.e., the Reynolds number of the flow varies drastically between the two cases, making the collapse of the curves less arbitrary.

Figure \ref{fig:DNS_spectra}(a) presents the dimensional energy spectra from the two simulations, whereas  figures \ref{fig:DNS_spectra}(b) and \ref{fig:DNS_spectra}(c) show the same data with the axes rescaled according to inertial and Kolmogorov similarity variables, respectively. The collapse of the large (small) scales is adequate only when the axes are normalized using the integral (Kolmogorov) scales, in agreement with a multiscale view of the cascade. However,  data from the same simulation size retain a `good-enough' collapse independent of the choice of normalization length, similar to our grid measurements, a fact which, if viewed in isolation, might give a false impression of a single length scale process. Figure \ref{fig:DNS_spectra}(d) shows the algorithmically identified collapse of the data, that is, using the empirical transformations
\begin{subequations}
\begin{equation}
\tilde{\kappa} \propto \kappa L^{0.342} \eta^{0.658}
\end{equation}
\begin{equation}
\tilde{E_{11}} \propto E_{11} \epsilon^{-0.655} L^{-0.561} \eta^{-1.094} k^{-0.017} \approx 
E_{11} \epsilon^{-0.667} \left( L^{0.344} \eta^{0.656} \right)^{-5/3}.
\end{equation}
\label{eq:dnsgoto}
\end{subequations}
The algorithm again identifies, approximately, the Taylor microscale ($\lambda = L^{1/3}\eta ^{2/3})$ as the appropriate length for the collapse of the turbulence dynamics, but this time it cannot be claimed that the Taylor scale collapses both large and small-scale dynamics - only that it is an (algorithmically) optimal compromise when a single length scale collapse is enforced. We conjecture that the reason behind this could be explained from Lundgren's two-scale turbulence theory which is based on the technique of matched asymptotic expansions \citep{lundgren2002kolmogorov,lundgren2003kolmogorov}. The repercussions of this theory are discussed in \cite{obligado2019non,meldi2021analysis}: in particular, it is suggested that the Taylor microscale is the length scale at which the cascade dynamics are optimally distanced from both non-equilibrium effects (large scales) and dissipative effects (small scales) at the same time (in a matched asymptotic manner). The algorithmically identified collapse (figure \ref{fig:DNS_spectra}(d)) produces best results in the intermediate, inertial range of the cascade where both above effects tend to become negligible. Based on this argument, the algorithm identifies the Taylor microscale scale because it is the optimal compromise of the two governing scales (inertial, Kolmogorov) of the cascade.

\section{Discussion and conclusion}
\label{sec:discussion}
We have presented a methodology for identifying similarity variables from data, in the absence of governing equations. Given measurements of a quantity of interest, the method computes the discrete values of the similarity transformations that best collapse the data and expresses their analytic form via symbolic regression. The self-similarity that we identify is the optimal  fit of the data, and is not derived from the governing partial differential equations and the boundary conditions of the specific problem. 

The proposed method differs from dimensionless learning approaches that aim to reduce the dimensionality of parameter spaces via identification of dimensionless groups \citep{xie2022data,bakarji2022dimensionally,yuan2025dimensionless} because our focus is on discovering the similarity transformations that enable the collapse of entire solution profiles. Our approach is not restricted to dilational similarity transformations, and is capable of identifying more general, non-uniform relationships between variables, as demonstrated in the Burgers’ equation example. This flexibility broadens the range of problems for which similarity solutions can be uncovered. In practical terms, this approach also reduces the degree of prior knowledge required to reveal self-similar behaviour. For example, in the turbulent wake case, methods based on dimensional analysis would require the characteristic velocity difference to be provided as input in order to reveal the self-similarity. On the other hand, our method can uncover it autonomously. More generally, Step 1 of the proposed framework can be used to detect self-similar behaviour whilst  being effectively agnostic to the underlying physics, i.e., without requiring prior specification of the relevant variables or parameters. In the symbolic regression step (Step 2), the user must supply the candidate parameters to obtain explicit analytic expressions, but the ability to first reveal the existence of self-similarity without specifying inputs may be particularly appealing for exploratory studies or complex systems. On the other hand, for cases of pure dilational self-similarity, methods based on dimensional analysis are expected to be more efficient. A note of caution: the computation of the $l_2$ norms in Eqn. \eqref{eq:optstep1} is performed over the common support of the transformed profiles. If the supports differ significantly, the optimization may become challenging. Although this issue did not arise in this work, we acknowledge it as a point that might need to be addressed in future work.

The capabilities of the method were demonstrated in five fluid mechanical problems, which are known or assumed to accept similarity solutions, based on data from both numerical and laboratory experiments. In four problems, the transformations identified by the method were in agreement to the analytically or empirically derived similarity transformations, whilst  circumventing the need for scaling analysis or similarity arguments. In the fifth problem (turbulence cascade) our method identified an empirical length scale as the optimal similarity length of the cascade, which was  close to the Taylor microscale. Further investigation revealed that this result cannot be treated as being unequivocally in support of single length scale theories of turbulence \citep{de1938statistical,sedov2018similarity,barenblatt1974theory,george1992decay}, but rather expresses the role of the Taylor microscale as an intermediate scale between the integral and Kolmogorov scales, at which non-equilibrium and dissipative effects are optimally spaced \citep{meldi2021analysis}.  

The results illustrate ways that the proposed algorithm can be used for the identification of similarity and scaling laws in situations where rigorous mathematical analysis is challenging. This includes a wide range of applications in fluid physics but also in other processes such as quasicrystal shape and growth \citep{kamiya2018discovery}, stellar collapse \citep{yahil1983self}, single protein dynamics \citep{hu2016dynamics}, and others. Other directions for future work include using the method to detect symmetry breaking, distinguish similarity across multiple scales, and identify the onset or breakdown of self-similarity.

%\backsection[Supplementary data]{\label{SupMat}Supplementary material and movies are available at \\https://doi.org/10.1017/jfm.2019...}

\backsection[Acknowledgements]{The authors would like to thank Elif Beko\u{g}lu for providing the flume data and the schematic of the experimental set-up. KS is grateful to Prof Susumu Goto for providing the periodic box DNS data.}

\backsection[Funding]{Parts of this work were conducted while NB was at Imperial College London, supported by EPSRC, Grant No. EP/W026686/1. LM acknowledges support from the  ERC Starting Grant PhyCo No. 949388 and the EU-PNRR YoungResearcher TWIN ERC-PI\_0000005. KS acknowledges support from the ERC Starting Grant ONSET No. 101163321.}

\backsection[Declaration of interests]{The authors report no conflict of interest.}

\backsection[Data availability statement]{The code and input data required to reproduce the work reported in the manuscript are available in the GitHub repository \url{https://github.com/nbeb/extracting_self-similarity_from_data}. The grid turbulence data are available upon request.}

\backsection[Author ORCIDs]{N. Bempedelis, https://orcid.org/0000-0002-7359-9144; L. Magri, https://orcid.org/0000-0002-0657-2611; K. Steiros, https://orcid.org/0000-0001-7779-170X}

\backsection[Author contributions]{NB, LM and KS designed research; NB and KS performed research; NB analyzed data; and NB and KS wrote the paper.}

\appendix
\section{Data-driven similarity inference: implementation example}
\label{app:algorithm}

A pseudocode exemplifying the implementation of the proposed method for the case of a quantity of interest $q(s,t)$ is provided in Algorithm \ref{ref:algo}. Clarifications or examples are shown in smaller font size below each pseudocode line. More implementation examples can be found in the code and input data that reproduce the problems considered in the paper. These are available in the GitHub repository \url{https://github.com/nbeb/extracting_self-similarity_from_data}.

\begin{algorithm}
\caption{Data-driven similarity inference: implementation example}
\begin{algorithmic}[1]
\STATE Gather a set of $n_t$ observations $q(s)$ \\
{\scriptsize{$\boldsymbol{s} \in \mathbb{R}^{n_s \times n_t}, \; \boldsymbol{q} \in \mathbb{R}^{n_s \times n_t}$ }}
\STATE Break the generalized similarity variable decomposition down to a set of admissible/relevant candidate transformations \\ 
{\scriptsize{e.g., pure dilation for $s$ ($\beta=0$), and pure ($\delta=0$) or generalised dilation ($\delta=\delta(t)$) for $q$. The generalized decomposition is $\xi = \alpha(t) s + \beta(s,t)$ and $\tilde{q} = \gamma(t) q + \delta(s,t)$}.}
\STATE \textit{// Step 1: Search for similarity variables}
\FOR{each combination $\in$ set of candidate transformations}
    \STATE Initialise design variable matrix $\boldsymbol{w}$ \\
    {\scriptsize{Dimensions of $\boldsymbol{w}$ depend on the nature of the candidate transformations. In the case of pure dilation for both $s$ and $q$, $\boldsymbol{w} \in \mathbb{R}^{n_t \times 2}$, with the first column corresponding to $s$ and the second column to $q$ (i.e., containing the discrete values of $\alpha$ and $\gamma$, respectively).}}
    \STATE \textit{// Optimization problem}
    \FOR{$\text{iter} = 1$ \TO \textrm{stopping criterion}}
      \STATE Apply transformations to observations \\
      {\scriptsize{$\boldsymbol{\xi} = \boldsymbol{s} \odot \left(\textrm{col}_1 \left(\boldsymbol{w} \right)\right)^\intercal, \boldsymbol{\tilde{q}} = \boldsymbol{q} \odot \left(\textrm{col}_2 \left(\boldsymbol{w} \right)\right)^\intercal$}}  
      \STATE Interpolate $\boldsymbol{\tilde{q}}$ on common $\boldsymbol{\xi}$ \\
      {\scriptsize{Resolution of interpolation grid may vary.}}
      \STATE Compute objective function (OF) (Eq. \eqref{eq:optstep1} or \eqref{eq:optstep1_var})
      \STATE Update $\boldsymbol{w}$ to minimize the OF
    \ENDFOR
    \IF{OF $\ll 1$}
    \STATE $\boldsymbol{w}_\textrm{opt} \leftarrow \boldsymbol{w}$
    \STATE Exit step 1: Similarity variables found. Potentially identified self-similarity.  
    \ENDIF
\ENDFOR
\STATE \textit{// Step 2: Analytic form of the transformations}   
\STATE Normalize transformations and variables found in step 1 (optional)
\FOR{each transformation component (i.e., column $j$ of $\boldsymbol{w}_\textrm{opt})$}
  \STATE Define the $n_v$ variables $\boldsymbol{A} \in \mathbb{R}^{n_v \times 1}$ that the component can be a function of
  \STATE Construct a matrix $\boldsymbol{A}_D$ containing the values of $\boldsymbol{A}$ at all locations \\
   {\scriptsize{In the example above, $\boldsymbol{A}_D \in \mathbb{R}^{n_v \times n_t}$}}
  \STATE \textit{// Symbolic regression task}   
  \STATE Find expression $\psi (\boldsymbol{A})$ that minimises \\ ${\| \psi(\boldsymbol{A}_D) - \textrm{col}_j \left(\boldsymbol{w}_{\textrm{opt}} \right) \|_2^2 + w_D \| \left[ \psi(\boldsymbol{A}) \right] - \left[ \textrm{col}_j \left(\boldsymbol{w}_{\textrm{opt}} \right) \right] \| }$ \\
  {\scriptsize{The brackets $[]$ denote the physical dimensions of a quantity in the form of a dimension vector, i.e., a vector containing the powers of its basic dimensions. For example, velocity has an associated dimension vector in the MLT (Mass, Length, Time) system $[u]=(0,1,-1)$. Note that the term which enforces dimensional homogeneity is optional and only relevant when the quantity of interest is dimensional.}}
\ENDFOR
\RETURN $\boldsymbol{w}, \psi(\boldsymbol{A})$
\end{algorithmic}
\label{ref:algo}
\end{algorithm}

\section{Sensitivity and robustness}
\label{app:sensitivity}
In this section, we present results from a sensitivity analysis with respect to the (i) number of provided profiles (stations), (ii) number of points at each station, (iii) number of points where the $l_2$ norms of Eqn. \eqref{eq:optstep1} are evaluated (i.e., discretisation of the transformed coordinates grid), and (iv) noise in the input data, for the Blasius boundary layer problem.

\begin{enumerate}
\item Number of provided stations: In the example detailed in section \ref{sec:results}, the algorithm is provided with velocity profiles at 9 different streamwise stations. Table \ref{tab:numberstations} shows the identified scalings when fewer stations at regular or irregular spacings are considered. The algorithm shows robust performance across all tested cases. 
\end{enumerate} 

\begin{table}
\begin{center}
\def~{\hphantom{0}}
\begin{tabular}{ccc}
Stations & $\tilde{y}/y$ & $\tilde{u}/u$ \\ \hline
$\left[1,2,3,4,5,6,7,8,9\right]$ & $U_\infty^{0.5018} \nu^{-0.5018} x^{-0.4981}$ & $U_\infty^{-0.9995} \nu^{-0.0005} x^{0.0005}$ \\ 
$\left[1,3,5,7,9\right]$ & $U_\infty^{0.5010} \nu^{-0.5010} x^{-0.4990}$ & $U_\infty^{-0.9980} \nu^{-0.0020} x^{0.0020}$ \\ 
$\left[1,3,9\right]$ & $U_\infty^{0.5008} \nu^{-0.5008} x^{-0.4992}$ & $U_\infty^{-0.9998} \nu^{-0.0002} x^{0.0002}$ \\ 
$\left[2,6,8\right]$ & $U_\infty^{0.5022} \nu^{-0.5022} x^{-0.4978}$ & $U_\infty^{-0.9990} \nu^{-0.0010} x^{0.0010}$ \\ %\hline
\end{tabular}
\caption{Data-driven identification of self-similarity in the Blasius boundary layer. Identified transformations for different number of available profiles (stations).}
\label{tab:numberstations}
\end{center}
\end{table}

\begin{enumerate}
\setcounter{enumi}{1}
\item Number of points at each station ($N_{meas}$): In the example detailed in section \ref{sec:results}, the velocity profiles at each streamwise station consist of 100 points. Table \ref{tab:numberpoints} shows the identified scalings when fewer points are available (50 and 20, respectively). The algorithm shows robust performance across all tested cases. A slight increase in the $y$ scaling error is observed as the number of available points decreases.
\end{enumerate}
\begin{table}
\centering
\begin{tabular}{ccc}
$N_{meas}$ & $\tilde{y}/y$ & $\tilde{u}/u$ \\ \hline
100 & $U_\infty^{0.5018} \nu^{-0.5018} x^{-0.4981}$ & $U_\infty^{-0.9995} \nu^{-0.0005} x^{0.0005}$ \\ 
50 & $U_\infty^{0.5033} \nu^{-0.5033} x^{-0.4967}$ & $U_\infty^{-0.9997} \nu^{-0.0003} x^{0.0003}$ \\ 
20 & $U_\infty^{0.5115} \nu^{-0.5115} x^{-0.4885}$ & $U_\infty^{-1.0001} \nu^{0.0001} x^{-0.0001}$ \\
\end{tabular}
\caption{Data-driven identification of self-similarity in the Blasius boundary layer. Identified transformations for different number of measurements available at each station.}
\label{tab:numberpoints}
\end{table}

\begin{enumerate}
\setcounter{enumi}{2}
\item Number of points in the transformed coordinates grid ($N_{\xi-points}$): As discussed in section \ref{sec:methodology}, the $l_2$-norms in Equation \eqref{eq:optstep1} are computed following interpolation on the transformed coordinates ($\xi$) grid. In the example detailed in section \ref{sec:results}, $N_{\xi-points}$ is set equal to the number of points in the input data $N_{meas}$. Table \ref{tab:interpgrid} shows the identified scalings when this ratio changes. The algorithm shows robust performance across all tested cases. 
\end{enumerate}
\begin{table}
\centering
\begin{tabular}{ccc}
$N_{\xi-points}/N_{meas}$ & $\tilde{y}/y$ & $\tilde{u}/u$ \\ \hline
0.25 & $U_\infty^{0.4993} \nu^{-0.4991} x^{-0.5013}$ & $U_\infty^{-0.9997} \nu^{-0.0003} x^{0.0003}$ \\
1.00 & $U_\infty^{0.5018} \nu^{-0.5018} x^{-0.4981}$ & $U_\infty^{-0.9995} \nu^{-0.0005} x^{0.0005}$ \\ 
4.00 & $U_\infty^{0.5016} \nu^{-0.5016} x^{-0.4984}$ & $U_\infty^{-0.9992} \nu^{-0.0008} x^{0.0008}$ \\
\end{tabular}
\caption{Data-driven identification of self-similarity in the Blasius boundary layer. Identified transformations for different discretisations of the transformed coordinates grid.}
\label{tab:interpgrid}
\end{table}

\begin{figure}
\centering
\includegraphics[width=\textwidth]{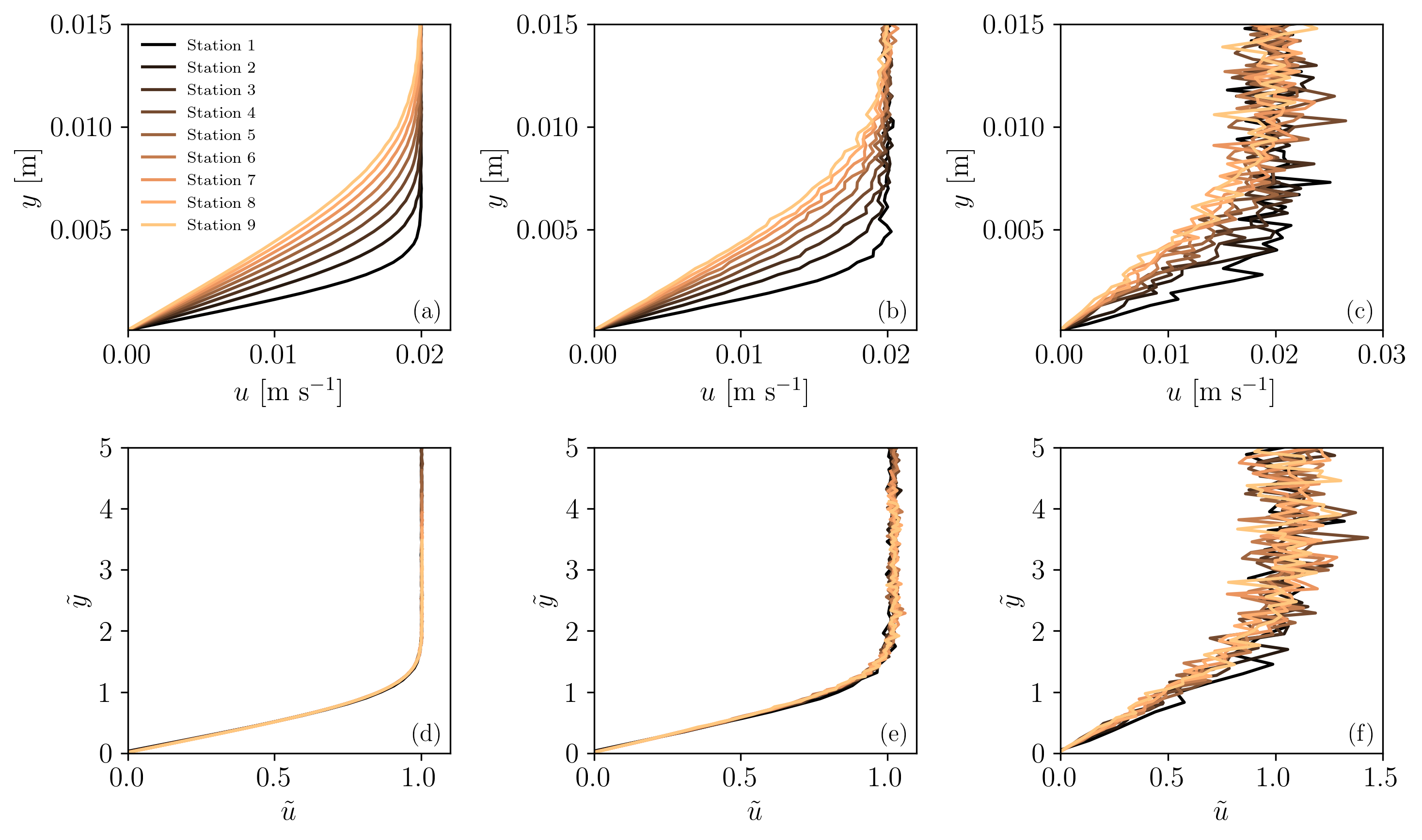}
\caption{Data-driven identification of self-similarity in the Blasius boundary layer with added noise. Top row: input data, bottom row: input data collapsed with the algorithmically identified scalings (table \ref{tab:noise}). (a,d) $\epsilon=0.001$, (b,e) $\epsilon=0.01$, (c,f) $\epsilon=0.1$.}
\label{fig:noise}
\end{figure}

\begin{enumerate}
\setcounter{enumi}{3}
\item Lastly, we consider the robustness of the algorithm to noise. To this end, we add zero-mean Gaussian noise to the data, with standard deviation proportional to the local velocity magnitude, scaled by a relative noise level $\epsilon$. The identified scalings for three different noise levels are given in table \ref{tab:noise}. The input and algorithmically scaled data are shown in figure \ref{fig:noise}. While the algorithm's accuracy decreases with increasing noise, the higher levels tested exceed typical experimental uncertainties, and the method can be expected to perform reliably under realistic noise conditions, as demonstrated in the examples using experimental datasets.
\end{enumerate}
\begin{table}
\centering
\begin{tabular}{ccc}
$\epsilon$ & $\tilde{y}/y$ & $\tilde{u}/u$ \\ \hline
0 & $U_\infty^{0.5018} \nu^{-0.5018} x^{-0.4981}$ & $U_\infty^{-0.9995} \nu^{-0.0005} x^{0.0005}$ \\ 
0.001 & $U_\infty^{0.5023} \nu^{-0.5023} x^{-0.4976}$ & $U_\infty^{-1.0000} \nu^{-0.0000} x^{0.0000}$ \\ 
0.01 & $U_\infty^{0.5366} \nu^{-0.5366} x^{-0.4634}$ & $U_\infty^{-0.9950} \nu^{-0.0050} x^{0.0050}$ \\
0.1 & $U_\infty^{0.6689} \nu^{-0.6689} x^{-0.3311}$ & $U_\infty^{-0.9820} \nu^{-0.0180} x^{0.0180}$ 
\end{tabular}
\caption{Data-driven identification of self-similarity in the Blasius boundary layer. Identified transformations for different levels of added noise.}
\label{tab:noise}
\end{table}

\section{Application to data from literature: turbulent wake}
\label{app:wake}
We attempt to recover the self-similar expressions for turbulent wakes using our method and experimental data from the literature. We consider the mean velocities at five stations in the wake of a $47\%$ solid slender plate, as measured by \citet{cimbala1988large}, in turbulent conditions ($\text{Re}=5000$) (see figures \ref{fig:wake}(a) and (b)). Figure \ref{fig:wake}(c) shows the collapse that is obtained via the proposed method, assuming transformations of the form $\tilde{y} = \alpha(x) y$ and $\tilde{u} = \beta(x) \bar{u} + \gamma(x)$. The identified transformations are regressed using $\texttt{PySR}$, with $\alpha = \alpha(x, y_{w}, y_{1/2})$ (with $y_w(x)$ defined as $\bar{u}(x,\pm y_{w}) = 0.99 U_\infty$), $\beta=\beta(U_\infty, \bar{u}_{\text{cntr}}, x, \nu)$ and $\gamma=\gamma(U_\infty, \bar{u}_{\text{cntr}}, x, \nu)$. The library of operators consists of the four basic mathematical operations ($+,-,\times,\div$). The interpreted similarity transformations are 
\begin{subequations}
\begin{equation}
\tilde{y}= \alpha y = \frac{1.2188}{y_{1/2}(x)} y
\end{equation}
\begin{equation}
\tilde{u}(x,\tilde{y}) = \beta\bar{u} + \gamma = \frac{1}{U_\infty - \bar{u}_\text{cntr}} \bar{u} - \frac{\bar{u}_\text{cntr}}{U_\infty - \bar{u}_\text{cntr}} = 1 - \tilde{\zeta}
\end{equation}
\end{subequations}
which match (ignoring the arbitrary multiplicative constant) the self-similar expressions for turbulent wakes found in the literature \citep{tennekes1972first, pope_2000, townsend1976structure}.

\begin{figure}
\centering
\includegraphics[width=\textwidth]{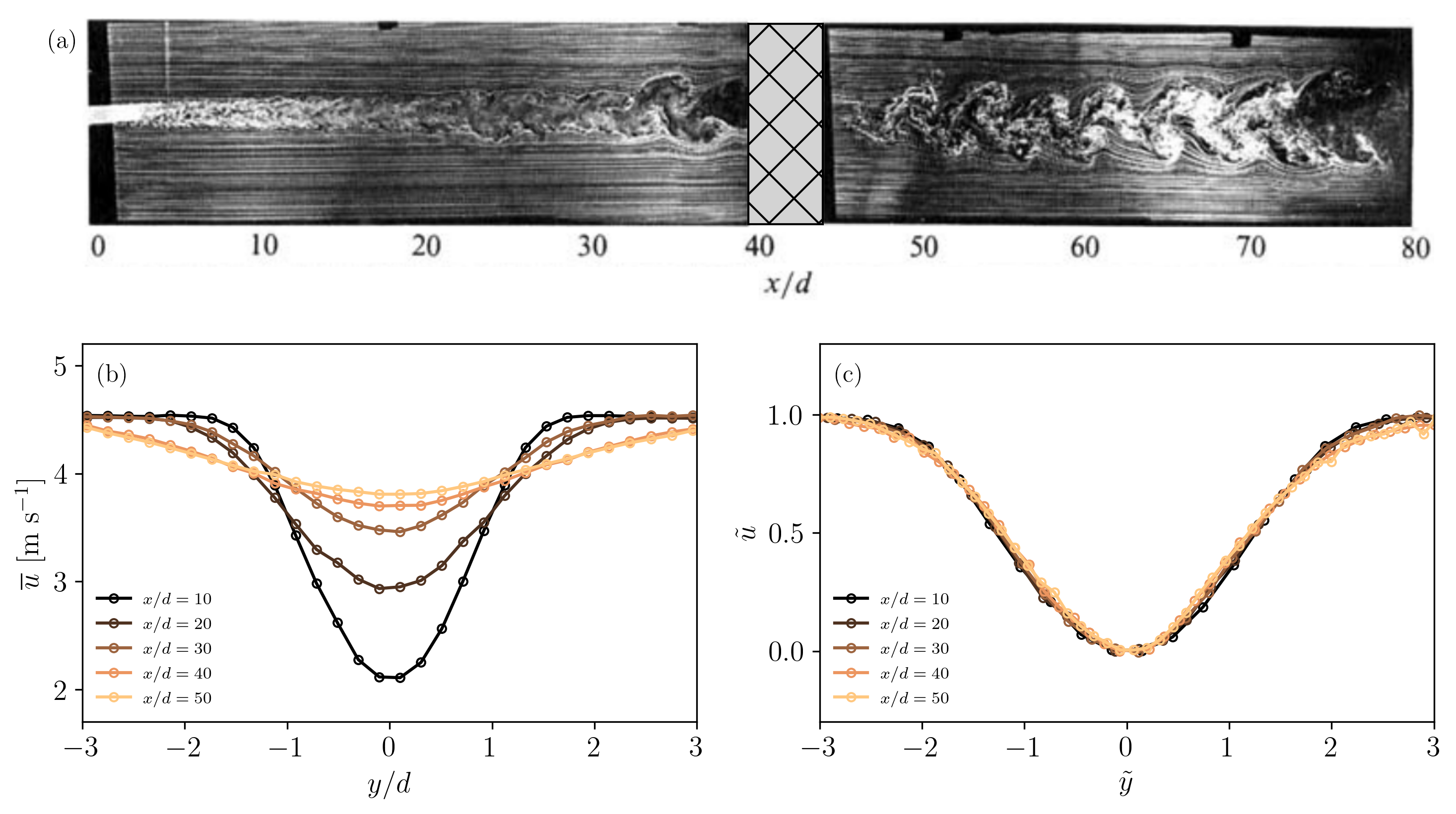}
\caption{Data-driven identification of self-similarity in the wake of a porous plate. (a) Smoke-wire visualisation of a porous plate wake. Figure adapted from \cite{cimbala1988large}. (b) Mean streamwise velocity profiles at different locations downstream of the plate. Experimental data extracted from \cite{cimbala1988large}. (c) Algorithmically collapsed velocity profiles.}
\label{fig:wake}
\end{figure}

\bibliographystyle{jfm}
\bibliography{references.bib}

\begin{thebibliography}{77}
\expandafter\ifx\csname natexlab\endcsname\relax\def\natexlab#1{#1}\fi
\def\au#1{#1} \def\ed#1{#1} \def\yr#1{#1}\def\at#1{#1}\def\jt#1{\textit{#1}}
  \def\bt#1{#1}\def\bvol#1{\textbf{#1}} \def\vol#1{#1} \def\pg#1{#1}
  \def\publ#1{#1}\def\arxiv#1{#1}\def\org#1{#1}\def\st#1{\textit{#1}}

\bibitem[Bakarji {\em et~al.\/}(2022)Bakarji, Callaham, Brunton \&
  Kutz]{bakarji2022dimensionally}
{\sc \au{Bakarji, J.}, \au{Callaham, J.}, \au{Brunton, S.~L.} \& \au{Kutz,
  J.~N.}} \yr{2022}  \at{{Dimensionally consistent learning with Buckingham
  Pi}}.  \jt{Nature Computational Science}  \bvol{2}~(12),  \pg{834--844}.

\bibitem[Barenblatt(1996)]{barenblatt1996scaling}
{\sc \au{Barenblatt, G.~I.}} \yr{1996} {\em Scaling, self-similarity, and
  intermediate asymptotics: dimensional analysis and intermediate
  asymptotics\/}.  \publ{Cambridge University Press}.

\bibitem[Barenblatt \& Gavrilov(1974)]{barenblatt1974theory}
{\sc \au{Barenblatt, G.~I.} \& \au{Gavrilov, A.~A.}} \yr{1974}  \at{On the
  theory of self-similar degeneracy of homogeneous isotropic turbulence}.
  \jt{Soviet Physics-JETP}  \bvol{38},  \pg{399--402}.

\bibitem[Batchelor(1953)]{batchelor1953theory}
{\sc \au{Batchelor, G.~K.}} \yr{1953} {\em The theory of homogeneous
  turbulence\/}.  \publ{Cambridge University Press}.

\bibitem[Beaumard {\em et~al.\/}(2024)Beaumard, Bragan{\c{c}}a, Cuvier, Steiros
  \& Vassilicos]{beaumard2024scale}
{\sc \au{Beaumard, P.}, \au{Bragan{\c{c}}a, P.}, \au{Cuvier, C.}, \au{Steiros,
  K.} \& \au{Vassilicos, J.~C.}} \yr{2024}  \at{{Scale-by-scale non-equilibrium
  with Kolmogorov-like scalings in non-homogeneous stationary turbulence}}.
  \jt{Journal of Fluid Mechanics}  \bvol{984},  \pg{A35}.

\bibitem[Bekoglu {\em et~al.\/}(2024)Bekoglu, Bempedelis \&
  Steiros]{ElifReview}
{\sc \au{Bekoglu, E.}, \au{Bempedelis, N.} \& \au{Steiros, K.}} \yr{2024} On
  the formation of the primary and secondary vortex street instabilities.
  \bt{In {\em 13th International Symposium on Turbulence and Shear Flow
  Phenomena (TSFP13)\/}}.

\bibitem[Bekoglu {\em et~al.\/}(2025)Bekoglu, Bempedelis \&
  Steiros]{bekogluSVS}
{\sc \au{Bekoglu, E.}, \au{Bempedelis, N.} \& \au{Steiros, K.}} \yr{2025}
  \at{Formation of turbulent secondary vortex street in absence of vortex
  shedding instability}.  \jt{Under review} .

\bibitem[Bempedelis \& Steiros(2022)]{bempedelis2022analytical}
{\sc \au{Bempedelis, N.} \& \au{Steiros, K.}} \yr{2022}  \at{Analytical
  all-induction state model for wind turbine wakes}.  \jt{Physical Review
  Fluids}  \bvol{7}~(3),  \pg{034605}.

\bibitem[Berny {\em et~al.\/}(2020)Berny, Deike, S{\'e}on \&
  Popinet]{berny2020role}
{\sc \au{Berny, A.}, \au{Deike, L.}, \au{S{\'e}on, T.} \& \au{Popinet, S.}}
  \yr{2020}  \at{Role of all jet drops in mass transfer from bursting bubbles}.
   \jt{Physical Review Fluids}  \bvol{5}~(3),  \pg{033605}.

\bibitem[Birkhoff(1960)]{birkhoff2015hydrodynamics}
{\sc \au{Birkhoff, G.}} \yr{1960} {\em Hydrodynamics\/}, ,  \vol{vol. 2234}.
  \publ{Princeton University Press}.

\bibitem[Blasius(1908)]{blasius1908}
{\sc \au{Blasius, H.}} \yr{1908}  \at{Grenzschichten in fl{\"u}ssigkeiten mit
  kleiner reibung}.  \jt{Zeitschrift f{\"u}r Mathematik und Physik}  \bvol{56},
   \pg{1--37}.

\bibitem[Cantwell(1978)]{cantwell1978similarity}
{\sc \au{Cantwell, B.~J.}} \yr{1978}  \at{Similarity transformations for the
  two-dimensional, unsteady, stream-function equation}.  \jt{Journal of Fluid
  Mechanics}  \bvol{85}~(2),  \pg{257--271}.

\bibitem[Cantwell(2002)]{cantwell2002introduction}
{\sc \au{Cantwell, B.~J.}} \yr{2002} {\em Introduction to symmetry analysis\/}.
   \publ{Cambridge University Press}.

\bibitem[Cimbala {\em et~al.\/}(1988)Cimbala, Nagib \&
  Roshko]{cimbala1988large}
{\sc \au{Cimbala, J.~M.}, \au{Nagib, H.~M.} \& \au{Roshko, A.}} \yr{1988}
  \at{Large structure in the far wakes of two-dimensional bluff bodies}.
  \jt{Journal of Fluid Mechanics}  \bvol{190},  \pg{265--298}.

\bibitem[Constantine {\em et~al.\/}(2017)Constantine, del Rosario \&
  Iaccarino]{constantine2017data}
{\sc \au{Constantine, P.~G.}, \au{del Rosario, Z.} \& \au{Iaccarino, G.}}
  \yr{2017}  \at{Data-driven dimensional analysis: algorithms for unique and
  relevant dimensionless groups}.  \jt{arXiv preprint arXiv:1708.04303} .

\bibitem[Cranmer(2023)]{cranmer2023interpretable}
{\sc \au{Cranmer, M.}} \yr{2023}  \at{{Interpretable machine learning for
  science with PySR and SymbolicRegression.jl}}.  \jt{arXiv preprint
  arXiv:2305.01582} .

\bibitem[Deike(2022)]{deike2022mass}
{\sc \au{Deike, L.}} \yr{2022}  \at{Mass transfer at the ocean--atmosphere
  interface: the role of wave breaking, droplets, and bubbles}.  \jt{Annual
  Review of Fluid Mechanics}  \bvol{54},  \pg{191--224}.

\bibitem[Deike {\em et~al.\/}(2018)Deike, Ghabache, Liger-Belair, Das, Zaleski,
  Popinet \& S{\'e}on]{deike2018dynamics}
{\sc \au{Deike, L.}, \au{Ghabache, E.}, \au{Liger-Belair, G.}, \au{Das, A.~K.},
  \au{Zaleski, S.}, \au{Popinet, S.} \& \au{S{\'e}on, T.}} \yr{2018}
  \at{Dynamics of jets produced by bursting bubbles}.  \jt{Physical Review
  Fluids}  \bvol{3}~(1),  \pg{013603}.

\bibitem[Desai {\em et~al.\/}(2022)Desai, Nachman \& Thaler]{desai2022symmetry}
{\sc \au{Desai, K.}, \au{Nachman, B.} \& \au{Thaler, J.}} \yr{2022}
  \at{Symmetry discovery with deep learning}.  \jt{Physical Review D}
  \bvol{105}~(9),  \pg{096031}.

\bibitem[Duchemin {\em et~al.\/}(2002)Duchemin, Popinet, Josserand \&
  Zaleski]{duchemin2002jet}
{\sc \au{Duchemin, L.}, \au{Popinet, S.}, \au{Josserand, C.} \& \au{Zaleski,
  Z.}} \yr{2002}  \at{Jet formation in bubbles bursting at a free surface}.
  \jt{Physics of Fluids}  \bvol{14}~(9),  \pg{3000--3008}.

\bibitem[Duraisamy {\em et~al.\/}(2025)Duraisamy, Brunton \&
  Taira]{duraisamy2025introduction}
{\sc \au{Duraisamy, K.}, \au{Brunton, S.~L.} \& \au{Taira, K.}} \yr{2025}
  \at{Introduction to turbulence \& learning from data}.  \bt{In {\em Data
  Driven Analysis and Modeling of Turbulent Flows\/}},  \pg{pp. 1--25}.
  \publ{Elsevier}.

\bibitem[Duraisamy {\em et~al.\/}(2019)Duraisamy, Iaccarino \&
  Xiao]{duraisamy2019turbulence}
{\sc \au{Duraisamy, K.}, \au{Iaccarino, G.} \& \au{Xiao, H.}} \yr{2019}
  \at{Turbulence modeling in the age of data}.  \jt{Annual Review of Fluid
  Mechanics}  \bvol{51},  \pg{357--377}.

\bibitem[Eggers(1997)]{eggers1997nonlinear}
{\sc \au{Eggers, J.}} \yr{1997}  \at{Nonlinear dynamics and breakup of
  free-surface flows}.  \jt{Reviews of Modern Physics}  \bvol{69}~(3),
  \pg{865}.

\bibitem[Eggers \& Fontelos(2015)]{eggers2015singularities}
{\sc \au{Eggers, J.} \& \au{Fontelos, M.~A.}} \yr{2015} {\em Singularities:
  formation, structure, and propagation\/}, ,  \vol{vol.~53}.  \publ{Cambridge
  University Press}.

\bibitem[Frisch(1995)]{frisch1995turbulence}
{\sc \au{Frisch, U.}} \yr{1995} {\em Turbulence: the legacy of AN
  Kolmogorov\/}.  \publ{Cambridge University Press}.

\bibitem[Fukami {\em et~al.\/}(2024)Fukami, Goto \& Taira]{fukami2024data}
{\sc \au{Fukami, K.}, \au{Goto, S.} \& \au{Taira, K.}} \yr{2024}
  \at{Data-driven nonlinear turbulent flow scaling with buckingham pi
  variables}.  \jt{Journal of Fluid Mechanics}  \bvol{984},  \pg{R4}.

\bibitem[Fukami \& Taira(2024)]{fukami2024single}
{\sc \au{Fukami, K.} \& \au{Taira, K.}} \yr{2024}  \at{Single-snapshot machine
  learning for super-resolution of turbulence}.  \jt{Journal of Fluid
  Mechanics}  \bvol{1001},  \pg{A32}.

\bibitem[Ga{\~n}{\'a}n-Calvo(2017)]{ganan2017revision}
{\sc \au{Ga{\~n}{\'a}n-Calvo, A.~M.}} \yr{2017}  \at{Revision of bubble
  bursting: Universal scaling laws of top jet drop size and speed}.
  \jt{Physical Review Letters}  \bvol{119}~(20),  \pg{204502}.

\bibitem[George(1989)]{george1989self}
{\sc \au{George, W.~K.}} \yr{1989}  \at{The self-preservation of turbulent
  flows and its relation to initial conditions and coherent structure}.
  \jt{Advances in Turbulence} .

\bibitem[George(1992)]{george1992decay}
{\sc \au{George, W.~K.}} \yr{1992}  \at{The decay of homogeneous isotropic
  turbulence}.  \jt{Physics of Fluids A: Fluid Dynamics}  \bvol{4}~(7),
  \pg{1492--1509}.

\bibitem[Ghabache {\em et~al.\/}(2014)Ghabache, Antkowiak, Josserand \&
  S{\'e}on]{ghabache2014physics}
{\sc \au{Ghabache, E.}, \au{Antkowiak, A.}, \au{Josserand, C.} \& \au{S{\'e}on,
  T.}} \yr{2014}  \at{On the physics of fizziness: How bubble bursting controls
  droplets ejection}.  \jt{Physics of Fluids}  \bvol{26}~(12).

\bibitem[Goto \& Vassilicos(2016)]{goto2016unsteady}
{\sc \au{Goto, S.} \& \au{Vassilicos, J.~C.}} \yr{2016}  \at{Unsteady
  turbulence cascades}.  \jt{Physical Review E}  \bvol{94}~(5),  \pg{053108}.

\bibitem[Gross(1996)]{gross1996role}
{\sc \au{Gross, D.~J.}} \yr{1996}  \at{The role of symmetry in fundamental
  physics}.  \jt{Proceedings of the National Academy of Sciences}
  \bvol{93}~(25),  \pg{14256--14259}.

\bibitem[Hu {\em et~al.\/}(2016)Hu, Hong, Dean~Smith, Neusius, Cheng \&
  Smith]{hu2016dynamics}
{\sc \au{Hu, X.}, \au{Hong, L.}, \au{Dean~Smith, M.}, \au{Neusius, T.},
  \au{Cheng, X.} \& \au{Smith, J.~C.}} \yr{2016}  \at{The dynamics of single
  protein molecules is non-equilibrium and self-similar over thirteen decades
  in time}.  \jt{Nature Physics}  \bvol{12}~(2),  \pg{171--174}.

\bibitem[Jofre {\em et~al.\/}(2020)Jofre, del Rosario \&
  Iaccarino]{jofre2020data}
{\sc \au{Jofre, L.}, \au{del Rosario, Z.~R.} \& \au{Iaccarino, G.}} \yr{2020}
  \at{Data-driven dimensional analysis of heat transfer in irradiated
  particle-laden turbulent flow}.  \jt{International Journal of Multiphase
  Flow}  \bvol{125},  \pg{103198}.

\bibitem[Kamiya {\em et~al.\/}(2018)Kamiya, Takeuchi, Kabeya, Wada, Ishimasa,
  Ochiai, Deguchi, Imura \& Sato]{kamiya2018discovery}
{\sc \au{Kamiya, K.}, \au{Takeuchi, T.}, \au{Kabeya, N.}, \au{Wada, N.},
  \au{Ishimasa, T.}, \au{Ochiai, A.}, \au{Deguchi, K.}, \au{Imura, K.} \&
  \au{Sato, N.~K.}} \yr{2018}  \at{Discovery of superconductivity in
  quasicrystal}.  \jt{Nature Communications}  \bvol{9}~(1),  \pg{154}.

\bibitem[von Karman \& Howarth(1938)]{de1938statistical}
{\sc \au{von Karman, T.} \& \au{Howarth, L.}} \yr{1938}  \at{On the statistical
  theory of isotropic turbulence}.  \jt{Proceedings of the Royal Society of
  London. Series A. Mathematical and Physical Sciences}  \bvol{164}~(917),
  \pg{192--215}.

\bibitem[Keller \& Miksis(1983)]{keller1983surface}
{\sc \au{Keller, J.~B.} \& \au{Miksis, M.~J.}} \yr{1983}  \at{Surface tension
  driven flows}.  \jt{SIAM Journal on Applied Mathematics}  \bvol{43}~(2),
  \pg{268--277}.

\bibitem[Kolmogorov(1941{\natexlab{{\em a\/}}})]{kolmogorov1941a}
{\sc \au{Kolmogorov, A.~N.}} \yr{1941{\natexlab{{\em a\/}}}} Dissipation of
  energy in the locally isotropic turbulence.  \bt{In {\em Dokl. Akad. Nauk
  SSSR A\/}}, ,  \vol{vol.~32},  \pg{pp. 16--18}.

\bibitem[Kolmogorov(1941{\natexlab{{\em b\/}}})]{kolmogorov1941c}
{\sc \au{Kolmogorov, A.~N.}} \yr{1941{\natexlab{{\em b\/}}}} The local
  structure of turbulence in incompressible viscous fluid for very large
  {R}eynolds numbers.  \bt{In {\em Dokl. Akad. Nauk SSSR A\/}}, ,
  \vol{vol.~30},  \pg{pp. 301--5}.

\bibitem[Kolmogorov(1941{\natexlab{{\em c\/}}})]{kolmogorov1941b}
{\sc \au{Kolmogorov, A.~N.}} \yr{1941{\natexlab{{\em c\/}}}} On degeneration
  (decay) of isotropic turbulence in an incompressible viscous fluid.  \bt{In
  {\em Dokl. Akad. Nauk SSSR A\/}}, ,  \vol{vol.~31},  \pg{pp. 538--40}.

\bibitem[Lai {\em et~al.\/}(2018)Lai, Eggers \& Deike]{lai2018bubble}
{\sc \au{Lai, C.-Y.}, \au{Eggers, J.} \& \au{Deike, L.}} \yr{2018}  \at{Bubble
  bursting: Universal cavity and jet profiles}.  \jt{Physical Review Letters}
  \bvol{121}~(14),  \pg{144501}.

\bibitem[Landau \& Lifshitz(1976)]{Landau1976}
{\sc \au{Landau, L.~D.} \& \au{Lifshitz, E.~M.}} \yr{1976} {\em Mechanics\/}.
  \publ{3rd ed. Pergamon Press}.

\bibitem[Liu \& Tegmark(2022)]{liu2022machine}
{\sc \au{Liu, Z.} \& \au{Tegmark, M.}} \yr{2022}  \at{Machine learning hidden
  symmetries}.  \jt{Physical Review Letters}  \bvol{128}~(18),  \pg{180201}.

\bibitem[Lundgren(2002)]{lundgren2002kolmogorov}
{\sc \au{Lundgren, Thomas~S}} \yr{2002}  \at{Kolmogorov two-thirds law by
  matched asymptotic expansion}.  \jt{Physics of fluids}  \bvol{14}~(2),
  \pg{638--642}.

\bibitem[Lundgren(2003)]{lundgren2003kolmogorov}
{\sc \au{Lundgren, T.~S.}} \yr{2003}  \at{Kolmogorov turbulence by matched
  asymptotic expansions}.  \jt{Physics of Fluids}  \bvol{15}~(4),
  \pg{1074--1081}.

\bibitem[Meldi \& Vassilicos(2021)]{meldi2021analysis}
{\sc \au{Meldi, M.} \& \au{Vassilicos, J.~C.}} \yr{2021}  \at{Analysis of
  lundgren's matched asymptotic expansion approach to the
  k{\'a}rm{\'a}n-howarth equation using the eddy damped quasinormal markovian
  turbulence closure}.  \jt{Physical Review Fluids}  \bvol{6}~(6),
  \pg{064602}.

\bibitem[Mendez \& Ordóñez(2004)]{mendez2004scaling}
{\sc \au{Mendez, P.~F.} \& \au{Ordóñez, F.}} \yr{2004}  \at{{Scaling Laws
  From Statistical Data and Dimensional Analysis}}.  \jt{Journal of Applied
  Mechanics}  \bvol{72}~(5),  \pg{648--657}.

\bibitem[Mototake(2023)]{mototake2023extracting}
{\sc \au{Mototake, Y.-i.}} \yr{2023} Extracting nonlinear symmetries from
  trained neural networks on dynamics data.  \bt{In {\em NeurIPS 2023 AI for
  Science Workshop\/}}.

\bibitem[Oberlack(1999)]{oberlack1999similarity}
{\sc \au{Oberlack, M.}} \yr{1999}  \at{Similarity in non-rotating and rotating
  turbulent pipe flows}.  \jt{Journal of Fluid Mechanics}  \bvol{379},
  \pg{1--22}.

\bibitem[Oberlack(2001)]{oberlack2001unified}
{\sc \au{Oberlack, M.}} \yr{2001}  \at{A unified approach for symmetries in
  plane parallel turbulent shear flows}.  \jt{Journal of Fluid Mechanics}
  \bvol{427},  \pg{299--328}.

\bibitem[Oberlack {\em et~al.\/}(2006)Oberlack, Cabot, Reif \&
  Weller]{oberlack2006group}
{\sc \au{Oberlack, M.}, \au{Cabot, W.}, \au{Reif, B. A.~Pettersson} \&
  \au{Weller, T.}} \yr{2006}  \at{Group analysis, direct numerical simulation
  and modelling of a turbulent channel flow with streamwise rotation}.
  \jt{Journal of Fluid Mechanics}  \bvol{562},  \pg{383--403}.

\bibitem[Oberlack {\em et~al.\/}(2022)Oberlack, Hoyas, Kraheberger,
  Alc{\'a}ntara-{\'A}vila \& Laux]{oberlack2022turbulence}
{\sc \au{Oberlack, M.}, \au{Hoyas, S.}, \au{Kraheberger, S.~V.},
  \au{Alc{\'a}ntara-{\'A}vila, F.} \& \au{Laux, J.}} \yr{2022}  \at{Turbulence
  statistics of arbitrary moments of wall-bounded shear flows: A symmetry
  approach}.  \jt{Physical Review Letters}  \bvol{128}~(2),  \pg{024502}.

\bibitem[Oberlack \& Rosteck(2010)]{oberlack2010new}
{\sc \au{Oberlack, M.} \& \au{Rosteck, A.}} \yr{2010}  \at{New statistical
  symmetries of the multi-point equations and its importance for turbulent
  scaling laws}.  \jt{Discrete Continuous Dyn. Syst}  \bvol{3},  \pg{451--471}.

\bibitem[Obligado \& Vassilicos(2019)]{obligado2019non}
{\sc \au{Obligado, M.} \& \au{Vassilicos, J.~C.}} \yr{2019}  \at{The
  non-equilibrium part of the inertial range in decaying homogeneous
  turbulence}.  \jt{Europhysics Letters}  \bvol{127}~(6),  \pg{64004}.

\bibitem[Otto {\em et~al.\/}(2023)Otto, Zolman, Kutz \&
  Brunton]{otto2023unified}
{\sc \au{Otto, S.~E.}, \au{Zolman, N.}, \au{Kutz, J.~N.} \& \au{Brunton,
  S.~L.}} \yr{2023}  \at{A unified framework to enforce, discover, and promote
  symmetry in machine learning}.  \jt{arXiv preprint arXiv:2311.00212} .

\bibitem[Pakdemirli \& Yurusoy(1998)]{pakdemirli1998similarity}
{\sc \au{Pakdemirli, M.} \& \au{Yurusoy, M.}} \yr{1998}  \at{Similarity
  transformations for partial differential equations}.  \jt{SIAM Review}
  \bvol{40}~(1),  \pg{96--101}.

\bibitem[Pope(2000)]{pope_2000}
{\sc \au{Pope, S.~B.}} \yr{2000} {\em Turbulent Flows\/}.  \publ{Cambridge
  University Press}.

\bibitem[Prandtl(1904)]{Prandtl904}
{\sc \au{Prandtl, L.}} \yr{1904} {\em Proc. Third Int. Math. Cong. Heidelberg
  1904\/}.

\bibitem[Saha {\em et~al.\/}(2021)Saha, Gan, Cheng, Gao, Kafka, Xie, Li,
  Tajdari, Kim \& Liu]{saha2021hierarchical}
{\sc \au{Saha, S.}, \au{Gan, Z.}, \au{Cheng, L.}, \au{Gao, J.}, \au{Kafka,
  O.~L.}, \au{Xie, X.}, \au{Li, H.}, \au{Tajdari, M.}, \au{Kim, H.~A.} \&
  \au{Liu, W.~K.}} \yr{2021}  \at{{Hierarchical deep learning neural network
  (HiDeNN): An artificial intelligence (AI) framework for computational science
  and engineering}}.  \jt{Computer Methods in Applied Mechanics and
  Engineering}  \bvol{373},  \pg{113452}.

\bibitem[Sanjay {\em et~al.\/}(2021)Sanjay, Lohse \&
  Jalaal]{sanjay2021bursting}
{\sc \au{Sanjay, V.}, \au{Lohse, D.} \& \au{Jalaal, M.}} \yr{2021}
  \at{Bursting bubble in a viscoplastic medium}.  \jt{Journal of Fluid
  Mechanics}  \bvol{922},  \pg{A2}.

\bibitem[Sedov(2018)]{sedov2018similarity}
{\sc \au{Sedov, L.~I.}} \yr{2018} {\em Similarity and dimensional methods in
  mechanics\/}.  \publ{CRC press}.

\bibitem[Steiros(2022{\natexlab{{\em a\/}}})]{steiros2022balanced}
{\sc \au{Steiros, K.}} \yr{2022{\natexlab{{\em a\/}}}}  \at{Balanced
  nonstationary turbulence}.  \jt{Physical Review E}  \bvol{105}~(3),
  \pg{035109}.

\bibitem[Steiros(2022{\natexlab{{\em b\/}}})]{steiros2022turbulence}
{\sc \au{Steiros, K.}} \yr{2022{\natexlab{{\em b\/}}}}  \at{Turbulence near
  initial conditions}.  \jt{Physical Review Fluids}  \bvol{7}~(10),
  \pg{104607}.

\bibitem[Tam(2019)]{tam2019phenomenological}
{\sc \au{Tam, C. K.~W.}} \yr{2019}  \at{A phenomenological approach to jet
  noise: the two-source model}.  \jt{Philosophical Transactions of the Royal
  Society A}  \bvol{377}~(2159),  \pg{20190078}.

\bibitem[Taylor(1950{\natexlab{{\em a\/}}})]{taylor1950formationA}
{\sc \au{Taylor, G.~I.}} \yr{1950{\natexlab{{\em a\/}}}}  \at{The formation of
  a blast wave by a very intense explosion i. theoretical discussion}.
  \jt{Proceedings of the Royal Society of London. Series A. Mathematical and
  Physical Sciences}  \bvol{201}~(1065),  \pg{159--174}.

\bibitem[Taylor(1950{\natexlab{{\em b\/}}})]{taylor1950formationB}
{\sc \au{Taylor, G.~I.}} \yr{1950{\natexlab{{\em b\/}}}}  \at{The formation of
  a blast wave by a very intense explosion.-ii. the atomic explosion of 1945}.
  \jt{Proceedings of the Royal Society of London. Series A. Mathematical and
  Physical Sciences}  \bvol{201}~(1065),  \pg{175--186}.

\bibitem[Tennekes \& Lumley(1972)]{tennekes1972first}
{\sc \au{Tennekes, H.} \& \au{Lumley, J.~L.}} \yr{1972} {\em A First Course in
  Turbulence\/}.  \publ{MIT Press}.

\bibitem[Townsend(1976)]{townsend1976structure}
{\sc \au{Townsend, A.~A.}} \yr{1976} {\em The structure of turbulent shear
  flow\/}.  \publ{Cambridge University Press}.

\bibitem[Vassilicos(2015)]{vassilicos2015dissipation}
{\sc \au{Vassilicos, J.~C.}} \yr{2015}  \at{Dissipation in turbulent flows}.
  \jt{Annual Review of Fluid Mechanics}  \bvol{47},  \pg{95--114}.

\bibitem[Whitham(2011)]{whitham2011linear}
{\sc \au{Whitham, G.~B.}} \yr{2011} {\em Linear and nonlinear waves\/}.
  \publ{John Wiley \& Sons}.

\bibitem[Xie {\em et~al.\/}(2022)Xie, Samaei, Guo, Liu \& Gan]{xie2022data}
{\sc \au{Xie, X.}, \au{Samaei, A.}, \au{Guo, J.}, \au{Liu, W.~K.} \& \au{Gan,
  Z.}} \yr{2022}  \at{Data-driven discovery of dimensionless numbers and
  governing laws from scarce measurements}.  \jt{Nature Communications}
  \bvol{13}~(1),  \pg{7562}.

\bibitem[Yahil(1983)]{yahil1983self}
{\sc \au{Yahil, A.}} \yr{1983}  \at{Self-similar stellar collapse}.  \jt{The
  Astrophysical Journal}  \bvol{265},  \pg{1047--1055}.

\bibitem[Yang {\em et~al.\/}(2023)Yang, Walters, Dehmamy \&
  Yu]{yang2023generative}
{\sc \au{Yang, J.}, \au{Walters, R.}, \au{Dehmamy, N.} \& \au{Yu, R.}}
  \yr{2023} Generative adversarial symmetry discovery.  \bt{In {\em
  International Conference on Machine Learning\/}},  \pg{pp. 39488--39508}.
  PMLR.

\bibitem[Yuan \& Lozano-Dur{\'a}n(2025)]{yuan2025dimensionless}
{\sc \au{Yuan, Y.} \& \au{Lozano-Dur{\'a}n, A.}} \yr{2025}  \at{Dimensionless
  learning based on information}.  \jt{arXiv preprint arXiv:2504.03927} .

\bibitem[Zeff {\em et~al.\/}(2000)Zeff, Kleber, Fineberg \&
  Lathrop]{zeff2000singularity}
{\sc \au{Zeff, B.~W.}, \au{Kleber, B.}, \au{Fineberg, J.} \& \au{Lathrop,
  D.~P.}} \yr{2000}  \at{Singularity dynamics in curvature collapse and jet
  eruption on a fluid surface}.  \jt{Nature}  \bvol{403}~(6768),
  \pg{401--404}.

\bibitem[Zel'Dovich \& Raizer(1967)]{zel1967physics}
{\sc \au{Zel'Dovich, Y.~B.} \& \au{Raizer, Y.~P.}} \yr{1967} {\em Physics of
  shock waves and high-temperature hydrodynamic phenomena\/}.  \publ{New York:
  Academic Press}.

\end{thebibliography}

\end{document}